\documentclass[fleqn,usenatbib]{mnras}

\usepackage{newtxtext,newtxmath}

\usepackage[T1]{fontenc}
\usepackage{ae,aecompl}
\usepackage{ulem}
\usepackage{float}

\usepackage{graphicx}	
\usepackage{amsmath}	
\usepackage{amssymb}	
\usepackage{color}
\definecolor{orange}{RGB}{210,105,30}

\title[Constraining the inner slope of massive clusters]{Constraining the inner density slope of massive galaxy clusters}

\author[QH~He et al.]{Qiuhan He$^{1,2}$\thanks{qiuhan.he@durham.ac.uk}, 
Hongyu Li$^{2}$,
Ran Li$^{2,3,4}$\thanks{liran827@gmail.com},
Carlos S. Frenk$^{1}$,
Matthieu Schaller$^{5}$,
\newauthor
David Barnes$^{6}$,
Yannick Bah\'e$^{5}$, 
Scott T. Kay$^{7}$, 
Liang Gao$^{1,2,3,4}$, 
Claudio Dalla Vecchia$^{8,9}$
\\
$^{1}$ Institute for Computational Cosmology, Department of Physics, University of Durham, South Road, Durham DH1 3LE, UK \\
$^{2}$ National Astronomical Observatories, Chinese Academy of Sciences, 20A Datun Road, Chaoyang District, Beijing 100012, China\\
$^{3}$ Key laboratory for Computational Astrophysics, National Astronomical Observatories, Chinese Academy of Sciences, Beijing, 100012, China\\
$^{4}$ School of Astronomy and Space Science, University of Chinese Academy of Science, 19A Yuquan Rd\\
$^{5}$ Leiden Observatory, Leiden University, PO Box 9513, 2300 RA Leiden, The Netherlands\\
$^{6}$ Department of Physics, Kavli Institute for Astrophysics and Space Research, Massachusetts Institute of Technology, Cambridge, MA 02139\\
USA\\
$^{7}$ Jodrell Bank Centre for Astrophysics, School of Physics and Astronomy, The University of Manchester, Manchester M13 9PL, UK\\
$^{8}$ Instituto de Astrof\'\i{}sica de Canarias, E-38205 La Laguna, Tenerife, Spain\\
$^{9}$ Universidad de La Laguna, Dpto. Astrof\'\i{}sica, E-38206 La Laguna, Tenerife, Spain\\
}

\date{}

\pubyear{}

\begin{document}
\bibliographystyle{mnras}
\pagerange{\pageref{firstpage}--\pageref{lastpage}}
\maketitle

\begin{abstract}
  We determine the inner density profiles of massive galaxy clusters
  (M$_{\rm 200}>5\times 10^{14}$~${\rm M}_{\odot}$) in the
  Cluster-EAGLE (C-EAGLE) hydrodynamic simulations, and investigate
  whether the dark matter density profiles can be correctly estimated
  from a combination of mock stellar kinematical and gravitational
  lensing data. From fitting mock stellar kinematics and lensing data
  generated from the simulations, we find that the inner density
  slopes of both the total and the dark matter mass distributions can
  be inferred reasonably well. We compare the density slopes of
  C-EAGLE clusters with those derived by Newman et al. for 7 massive
  galaxy clusters in the local Universe. We find that the asymptotic
  best-fit inner slopes of ``generalized'' NFW (gNFW) profiles,
  $\gamma_{\rm gNFW}$, of the dark matter haloes of the C-EAGLE
  clusters are significantly steeper than those inferred by Newman et
  al. However, the mean mass-weighted dark matter density slopes of
  the simulated clusters are in good agreement with the Newman et
  al. estimates. We also find that the estimate of
    $\gamma_{\rm gNFW}$ is very sensitive to the constraints from weak lensing
    measurements in the outer parts of the cluster and a bias can lead
    to an underestimate of $\gamma_{\rm gNFW}$.
\end{abstract}



\section{Introduction}

In the $\Lambda$CDM cosmological model cold dark matter dominates the
matter budget of the Universe, and much of it clusters into dark
matter halos. Gas condenses at the centres of these haloes, forming stars and giving birth to galaxies
\citep{White_Rees1978,White_Frenk1991}. Measuring the distributions of
dark and baryonic matter at the centres of haloes provides a key test
of $\Lambda$CDM and theories of galaxy formation.

Over the past three decades, the evolution of pure cold dark matter
has been calculated with great precision by means of N-body
simulations
\citep{DEFW,NFW96,NFW97,Jenkins2001,Diemand2007,Springel2008,Gao2011} 
\cite[for a review see][]{Frenk_White_2012}. In particular,
\citet[][hereafter NFW]{NFW96,NFW97} have shown that dark matter haloes have a
universal, self-similar, spherically averaged mass profile with
asymptotic behaviour, $\rho(r) \propto r^{-1}$, at the centre, and
$\rho(r) \propto r^{-3}$ at large radii.

In reality, in a bright galaxy baryonic matter dominates the mass
budget at the centre of the halo \citep{Schaller2015a}. Furthermore, the galaxy formation
process may modify the central halo density itself. The effects of these
baryonic processes are complex and even their sign is unclear: while
baryon condensation and contraction may sharpen the density profile
\citep{Blumenthal1986, Gnedin2004, Gustafsson2006,
  Duffy2010,Schaller2015a, Peirani2017}, rapid expulsion of gas due to feedback
process may flatten it, at least in faint galaxies
\citep[e.g.][]{NEF1996,
  Dehnen2005,Read2005,Mashchenko2006,Pontzen2012}. The
competition between these processes is best followed with
hydrodynamical simulations, but even then discrepancies persist. For
example, \citet{Gnedin2011} and \citet{Schaller2015a, Schaller2015b} show that the net effect of baryonic processes in
large galaxies in the field and in clusters is to preserve the
asymptotic dark matter density profile, $\rho(r)\propto -1$, but
\cite{Martizzi2013} find that cores may be generated by AGN feedback
in extreme cases.

Observationally, the inner density slopes of bright galaxies are best
constrained by combining stellar dynamics data for the central galaxy
with gravitational lensing data at large radii
\citep[e.g.][]{Treu2002, Treu2004, Auger2010,
  Sonnenfeld2015,Newman2013a,Newman2013b,Newman2015,Shu2015}. In this
way the {\it total} density profile of a galaxy can be measured, from
several kiloparsecs to tens of kiloparsecs from the centre. The total mass-averaged
density slope, $\bar{\gamma}$, within the effective radius of early
type galaxies is found to be around -2 in galaxy and group scale
systems, but may drop gradually to -1.2 in massive clusters
\citep{Treu2004, Auger2010, Newman2015, LiRan2019}. The dark matter halo profile
is not directly measurable and can only be inferred by assuming a
model to subtract the contribution from the stellar component. Recent
measurements have concluded that while the halo density profile in
groups is consistent with the NFW form \citep{Newman2015, Smith2017}, some
clusters the inner slope is around -0.5, significantly shallower than
the NFW prediction \citep{Sand2004, Sand2008, Newman2013b, Popolo2018}, and in contradiction with
cosmological simulation results.

There are several possible interpretations for this discrepancy. The
simulations may lack the correct physics, or treat baryonic processes
improperly \citep{Laporte2012, Laporte2015}, or it may be that the dark matter is not cold but perhaps
made up of self-interacting particles \citep[e.g.][]{Spergel2000,
  Vogelsberger2012, Rocha2013,Kaplinghat2016,Robertson2017a,Robertson2017b}. An
alternative explanation is that systematic effects in the analysis of
the observational data have been underestimated.

There are several potential sources of systematic uncertainties when
subtracting the stellar component in order to infer the inner slope of
the dark matter component. For example, the shape of the stellar
density profile is usually inferred from the light profile either
assuming a constant mass-to-light ratio or a stellar population
synthesis model
\citep[e.g.][]{Cappellari2008,Newman2013a,Newman2015}. A systematic
overestimation of the mass-to-light ratio could relieve the tension
between observations and theory \citep{Schaller2015b}. In addition,
simplistic assumptions about the symmetry of the system or the
anisotropy of the velocity distribution may also bias the inference of
the inner dark matter profile
\citep[e.g.][]{Meneghetti2007,Hongyu2016}.  Recently,
  \citet{Sartoris2020} have estimated a value of -0.99 for the dark
  matter density slope at the centre of Abell S1063 -- in excellent
  agreement with $\Lambda$CDM predictions -- from analysis of a 
  large sample of stars in the central galaxy with spectroscopic data
  and a model allowing for variable velocity dispersion anisotropy.

In this work we assess dark matter density reconstruction methods in
galaxy clusters that combine stellar dynamics with gravitational
lensing. We construct mock data using the C-EAGLE simulations, a set
of high resolution zoom-in hydrodynamical simulations of massive
clusters \citep{Barnes2017,Bahe2017}.  We then perform a combined
analysis of stellar kinematics and gravitational lensing on the mock
data and explore the accuracy of the recovered dark matter density
profiles.

The structure of the paper is as follows. In Section~\ref{sec:mock} we
describe our mock data and in Section~\ref{sec:dynamics} our models,
and the method used to infer model parameters. In
Section~\ref{sec:results} we present the recovery of dark matter
profiles and study the model dependence on galaxy shape and velocity
anisotropy. We summarize and discuss our results in
Section~\ref{sec:sum}.

\section{Mock data}
\label{sec:mock}
\subsection{The C-EAGLE simulations}

We create mock observations using the C-EAGLE simulations
\citep{Bahe2017, Barnes2017}. This set of cosmological hydrodynamical
simulations consists of 30 zoom-in resimulated massive galaxy clusters
that were selected from a larger volume dark matter-only simulation
according to a criterion based on halo mass and isolation
\citep{Bahe2017}. The C-EAGLE simulations employ the state-of-the-art
EAGLE galaxy formation model and simulation code \citep{Schaye2015, Crain2015}. This code is based
on a modified version of the \textsc{gadget-3} smooth particle
hydrodynamics (SPH) code last described in \citet{Springel2005}, which
include radiative cooling, star formation, stellar and black hole
feedback, etc. The parameters of the subgrid models used for EAGLE
were calibrated so as to reproduce a small subset of data of the z=0 field
galaxy population\citep{Schaye2015, Crain2015}. C-EAGLE made use of the AGNdT9
model which gives a better match than the reference EAGLE model to the
X-ray luminosities and gas fractions of low-mass galaxy groups
\citep{Schaye2015}. C-EAGLE adopted the same $\Lambda$CDM cosmological
parameters as EAGLE: H$_0=67.77$km s$^{-1}$ Mpc$^{-1}$,
$\Omega_{\Lambda}=0.693$, $\Omega_{\rm M}=0.307$ and
$\Omega_{\rm b}=0.04825$. The mass resolution of C-EAGLE is the same
as in EAGLE: $1.8\times10^6\,\mathrm{M}_{\odot}$ initially for gas
particles and $9.7\times10^6\,\mathrm{M}_{\odot}$ for dark matter
particles. The Plummer gravitational softening length of the
high-resolution region was set to 2.66 comoving $\mathrm{kpc}$ for
$z > 2.8$, and then kept fixed at $0.70$ physical $\mathrm{kpc}$ for
$z < 2.8$. The minimum smoothing length of the SPH kernel was set to a
tenth of the gravitational softening scale.

In this paper we are interested in massive clusters comparable to
those in the sample of \citet{Newman2013a, Newman2013b} and so we
focus on clusters whose mass falls in the range
$4.0\times10^{14}<M_{200}<2\times10^{15}\,\mathrm{M}_{\odot}$ at
$z=0$, where $M_{200}$ is the mass enclosed within a sphere of radius
$r_{200}$ whose mean density is 200 times the critical density of the
universe.  Altogether our sample consists of 17 massive galaxy
clusters, denoted by CE-12 to CE-28 in the C-EAGLE simulations
\citep{Barnes2017}. While the clusters analyzed by
  \citet{Newman2013a, Newman2013b} have an average redshift of $z\sim$
  0.2, the simulation output we analyze is at $z = 0$. However, we
  have checked that our conclusions are unaffected by this
  choice. Further properties of our clusters may be found in the
  tables in the Appendix of \citet{Barnes2017, Bahe2017}.

\subsection{Photometric and kinematic data}
We create photometric and kinematic mock data following a similar
process to that described by \citet{Hongyu2016}. First, we define the
central galaxy as the one lying closest to the centre of the potential
of the cluster. Using the same method as
  \citet{Schaller2015c}, we find that all our central galaxies are
  very close to the centre of the potential, with a mean offset of
  0.2~ kpc and a maximum of 0.8~kpc. Since the offset is comparable to
  the softening length of the simulations (0.70~kpc), the centres of
  the central galaxies are consistent with the centres of the
  potential. Next, we construct the surface stellar mass density map
of the central galaxies in the C-EAGLE clusters by projecting the
galaxy's star particles onto the $x-y$ plane of the
  simulation volume on a grid of cellsize  $0.5\times0.5\,\mathrm{kpc}^{2}$.

To generate a brightness map we assume a constant $M^*/L$ ratio for
each star particle. For comparison, we also generate a surface brightness map
for each central galaxy by calculating the mass-weighted r-band
brightness in each cell. The luminosities of individual
  star particles are derived following the method of
  \citet{Trayford2015}.

We then calculate the mean and standard deviation of the line-of-sight
velocities of stars in each cell. As \citet{Newman2013a}, we obtain
kinematic data in a long slit of width $3\,\mathrm{kpc}$ aligned with
the major axis of the galaxy. The bins extend from the galactic centre
to $21\,\mathrm{kpc}$, which is approximately $1.5$ effective radii
($R_E$) for the galaxies in our sample. We assume that
  the uncertainty in the measured velocity dispersion is $6$ percent
  in the inner four bins and $9$ percent in the outer three bins,
  similar to the values in \citet{Newman2013a}. For the situation
  where a satellite happens to lie along the line-of-sight, we discard
  the affected bins.

\begin{figure}
 \includegraphics[width=0.52\textwidth]{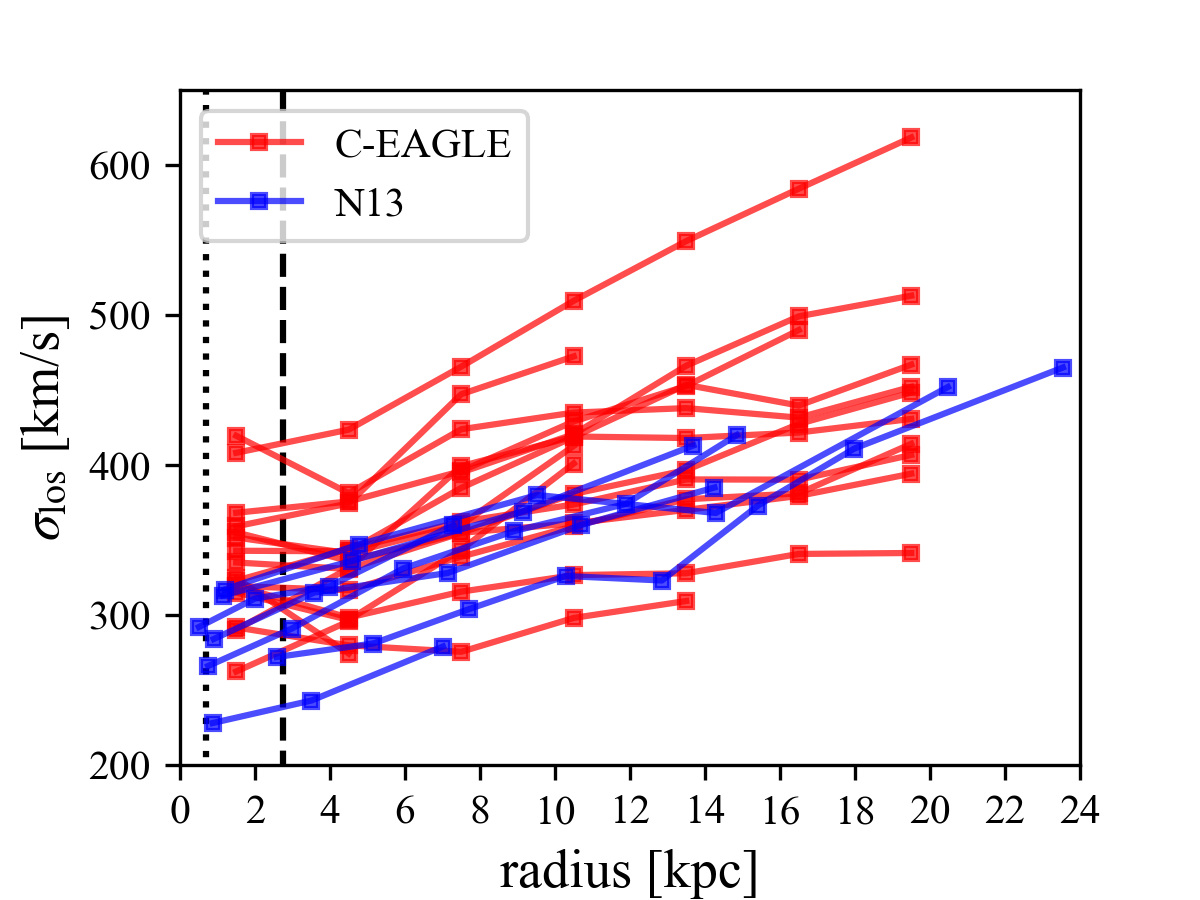}
 \caption{Line-of-sight velocity dispersion profiles. Red and blue
   lines represent profiles derived for our sample of BCGs and those
   from derived from the observed data in \citet{Newman2013a},
   respectively. The vertical dotted line marks the softening length and
the vertical dashed line the 3D average Power {\it et al.} radius for our
   clusters.}
 \label{fig:rms_compare}
 \end{figure}

In Fig.~\ref{fig:rms_compare}, we compare the line-of-sight velocity
dispersion profiles for our sample of clusters with those from
\citet{Newman2013a}. The blue points are the velocity dispersions of
the \citet{Newman2013a} clusters and the red points are those of our
clusters. The vertical dotted line marks the softening length and the vertical dashed line is the 3D average Power {\it et
    al.} radius \citep{Power2003}, which is usually taken to define
  the region where the profiles are numerically converged. Here, we
adopt the same threshold as \citet{Schaller2015a} to derive the Power
{\it et al.}  radius for our clusters. As we can see,
  most our clusters have higher line-of-sight velocity dispersions
  than the observed clusters. This is because at a given halo
  mass, the brightest cluster galaxies (BCGs) in C-EAGLE contain more
  stellar mass than observed BCGs by up to 0.6 dex \citep{Bahe2017}
  and this results in a greater mass concentration and thus a
  larger velocity dispersion reflecting the deeper gravitational
  potential.

 \subsection{Gravitational lensing mock data}\label{sec:gd}
 We calculate the tangential shear of the clusters at ten
   equally spaced logarithmic bins in radius ranging from
   $100\,\mathrm{kpc}$ to $2000\,\mathrm{kpc}$, which is similar to
   the range covered by the data of
   \citet{Newman2013a,Newman2013b}. Since the shear contributed by the
   correlation between different haloes is much smaller than the shear
   caused by the halo itself \citep{Cacciato2009, Li2009}, for
   simplicity, the lensing signal 
   is calculated only from the mass distribution in the halo ignoring
   the contribution of the large-scale structure. The tangential
 shear, $\gamma_t$, at projected radius, $R$, can be written as,

\begin{equation}\label{equ:weak}
\gamma_{t}(R)\Sigma_{\rm crit} =\Delta\Sigma(R)= M_{\rm surf}(R)/(\pi R^2) - \Sigma(R),
\end{equation}
where $M_{\rm surf}(R)$ is the mass, including dark
  matter, stars and gas, enclosed within projected radius, $R$;
$\Sigma(R)$ is the surface density at $R$; and $\Sigma_{\rm crit}$ is
the critical surface density, which is determined from the redshifts
of the lens and the source. We assume that the error on the tangential
shear is $40\%$, comparable to the average error in Fig.~5 of
\citet{Newman2013a}. In this work, we do not perturb the true value,
so the centre points of our weak lensing ``measurements'' are not
biased.

Strong lensing is also important in constraining the mass
  model of the clusters. The observations usually consist of multiple
  arcs produced by different background galaxies at different
  redshifts and those arcs can be very sensitive to the local surface
  density. The strong lensing constraint in \citet{Newman2013a} comes
  from measurements of the positions of multiple images, whose
  uncertainty is taken to be 0.5~ arcsecs. Here, to simplify our
  modelling, we approximate the strong lensing constraint as an
  aperture mass. The average Einstein radius, $R_{\rm Ein}$, of the
  Newman et al clusters is $\sim$ 10~arcsecs; thus we assume that the
  total projected mass within $R_{\rm Ein}$ ($\sim 39$~kpc at $z=0.2$)
  can be measured to a precision of $5\%$.

\section{Models}
\label{sec:dynamics}


We use two approaches to model the stellar kinematics of the central
galaxies in the C-EAGLE clusters:
\begin{enumerate}
\item the spherically symmetric Jeans model (sJ) 
\citep{Binney2008, Cappellari2008} 
\item the Jeans anisotropic model (JAM)
\citep{Emsellem1994,Cappellari2002,Cappellari2008}. 
\end{enumerate}
Since \citet{Newman2013a, Newman2013b} mainly used sJ, when comparing
our results with observations, we will mostly rely on this model as well. However, for an
interesting theoretical test, we also combine JAM with the lensing
analysis to investigate if it results in significant differences.

\subsection{sJ Model} 
For the spherically symmetric case, the Jeans equation gives the
relation between the line-of-sight velocity dispersion,
$\sigma_{\rm los}$, and the mass distribution, $M_{\rm tot}(r)$, as:
\begin{equation}\label{equ:sJ}
\sigma_{\rm los}(R) = \frac{2G}{\Sigma_*(R)}\int_{R}^{\infty}\frac{\rho_*(r)M_{\rm tot}(r)F(r)}{r^2}dr \:,
\end{equation}
where $\Sigma_*$ and $\rho_*$ are the surface density and 3D density
of the stars, respectively, $M_{\rm tot}(r)$ is the total mass
enclosed within 3D radius $r$ and, in the isotropic case, $F(r)=\sqrt{r^2-R^2}$. 

Following \citet{Newman2013a}, we use a 3-parameter dPIE model
\citep{Eliasdottir2007} to describe the 3D density profile of the
stellar component, where, 
\begin{equation}\label{equ:dPIE}
\rho_{\rm dPIE}(r) = \frac{\rho_0}{(1+r^2/a^2)(1+r^2/s^2)} \:;
\end{equation}
the core radius, $a$, the scale radius, $s$ ($s > a$), and the central
density, $\rho_0$, are free parameters. The surface density profile of
the stellar component can be analytically written as:
\begin{equation}\label{equ:dPIEs}
 \Sigma_{\rm dPIE}(R) = \rho_{0}\frac{\pi a^2s^2}{s^2-a^2}\left(\frac{1}{\sqrt{a^2+R^2}}-\frac{1}{\sqrt{s^2+R^2}}\right)\:.
\end{equation}
We fix the profile parameters, $a$ and $s$, by fitting the dPIE model to
the mass surface density profile of the central galaxy. Only the
normalization of the density profile is allowed to vary during the
dynamical modeling process.  

The mass distribution of the dark matter halo follows a gNFW profile:
\begin{equation}\label{equ:gNFW}
 \rho_{\rm gNFW}(r) = \rho_s\left(\frac{r}{r_s}\right)^{-\gamma_{\rm gNFW}}\left(\frac{1}{2}+\frac{1}{2}\frac{r}{r_s}\right)^{\gamma_{\rm gNFW}-3}\:,
\end{equation}
where $\rho_s$ is the characteristic density and $\gamma_{\rm gNFW}$ gives the inner asymptotic density
slope of the halo. For the NFW profile, $\gamma_{\rm gNFW}=1$.

For the spherical Jeans model we therefore have the following free parameters:
\begin{enumerate}

\item the stellar mass-to-light ratio: $M^*/L$;
\item the three parameters that describe the dark matter halo density profile:
  $\rho_{\rm s}$, $r_{s}$ and $\gamma_{\rm gNFW}$.
\end{enumerate}

\subsection{The JAM method}

For many galaxies in the real Universe, the assumption of spherical
symmetry for the distributions of mass and velocity dispersion are not
valid. In practice, assuming an axisymmetric mass distribution
often provides a better solution to galactic dynamical modeling.

For a steady-state axisymmetric mass distribution, the Jeans equations
in cylindrical coordinates, $(R,z,\phi)$, can be written as: 
\begin{eqnarray}
    \frac{n\overline{v_R^2}-n\overline{v_\phi^2}}{R}
    + \frac{\partial(n\overline{v_R^2})}{\partial R}
    + \frac{\partial(n\overline{v_R v_z})}{\partial z}
    & = & -n\frac{\partial\Phi_{\rm tot}}{\partial R},
    \label{eq:jeans_cyl_R}\\
    \frac{n\overline{v_R v_z}}{R}
    + \frac{\partial(n\overline{v_z^2})}{\partial z}
    + \frac{\partial(n\overline{v_R v_z})}{\partial R}
    & = & -n\frac{\partial\Phi_{\rm tot}}{\partial z},
    \label{eq:jeans_cyl_z}
\end{eqnarray}
where the $v$s denote the three components of velocity, 
\begin{equation}
    n\overline{v_k v_j}\equiv\int v_k v_j f\; \mathrm{d}^3 \mathbf{v},
\end{equation}
$f$ is the distribution function of the stars, $\Phi_{\rm tot}$
the gravitational potential, and $n$ is the luminosity density. 

In this work, we adopt the numerical Jeans-Anisotropic-Modeling
routine of \cite{Cappellari2008} with the multi-Gaussian Expansion
(MGE) technique \citep{Emsellem1994,Cappellari2002}, which is
widely used in galactic dynamical modeling
\citep[e.g.][]{Cappellari2008,Cappellari2011, Newman2015, Hongyu2016,
  Hongyu2017}

To determine a unique solution, the JAM routines make two assumptions
\citep{Cappellari2008}: 
\begin{enumerate}
\item the velocity dispersion ellipsoid is
aligned with the cylindrical coordinate system
($\overline{v_R v_z}=0$),

\item the anisotropy in the meridional plane is constant,
  i.e. $\overline{v_R^2} = b\overline{v_z^2}$, where $b$ is related to
  $\beta_{z}$, the anisotropy parameter in the $z$ direction, defined
  as

\begin{equation}
  \label{eq:beta}
  \beta_{z} \equiv 1 - \frac{\overline{v_z^2}}{\overline{v_R^2}} \equiv 1 - \frac{1}{b} \, .
\end{equation}

\end{enumerate}

If we set the boundary condition,  $n\overline{v_z^2}=0$ as $z\rightarrow\infty$,
the solution of Jeans equations can be written as  
\begin{eqnarray}
    n\overline{v_z^2}(R,z)
    & =  & \int_z^\infty n\frac{\partial\Phi_{\rm tot}}{\partial z}{\mathrm{d}} z
    \label{eq:jeans_sol_z}\\
    n\overline{v_\phi^2}(R,z) & = &
    b\left[
    R \frac{\partial(n\overline{v_z^2})}{\partial R}
    + n\overline{v_z^2} \right]
    + R n\frac{\partial\Phi_{\rm tot}}{\partial R}
    \label{eq:jeans_sol_R}.
\end{eqnarray}
The intrinsic velocity dispersions on the left-hand side of these equations
are integrated along the line-of-sight to derive the projected second
velocity moment, $\overline{v^{2}_{\mathrm{los}}}$. This can be
directly compared with the kinematical data for the stellar component,
i.e. the rms velocity,
$v_{\mathrm{rms}} \equiv \sqrt{v^{2} + \sigma^{2}}$, where $v$ and
$\sigma$ are the stellar mass-weighted line-of-sight velocity and
velocity dispersion, respectively. 

The gravitational potential, $\Phi_{\rm tot}$, is determined by the
total mass distribution. We consider two components: the stars and the
dark matter haloes. To speed up the calculation, the JAM routines use
Multi-Gaussian-Expansion \citep[MGE;][]{Emsellem1994} to fit the
surface brightness distribution, $\Sigma(x', y')$, of the central galaxies
\begin{equation}\label{equ:mge}
\Sigma(x', y') = \sum_{k=1}^{N}\frac{L_k}{2\pi\Delta_k^2q_k^{'2}}\text{exp}\left[-\frac{1}{2\Delta_k^2}\left(x^{'2}+\frac{y^{'2}}{q_k^{'2}}\right)\right]\:,
\end{equation}
where $L_k$ is the total luminosity of the $k$-th Gaussian component
with dispersion, $\Delta_k$, along the major axis, and $q'_k$ is the
projected axial ratio in the range [0,1]. The JAM routines assume
galaxies to be oblate axisymmetric. Thus, once the inclination angle
$i$ (i = 90$^\circ$ for edge-on) is known, the three dimensional
luminosity profile in cylindrical coordinates, $n(R, z)$, is given
by 
\begin{equation}\label{mge:mu}
n(R,z)= \sum_{k=1}^{N}\frac{L_k}{(\sqrt{2\pi\Delta_k})^3q_k}\text{exp}\left[-\frac{1}{2\Delta_k^2}\left(R^2+\frac{z^2}{q_k^2}\right)\right]\:.
\end{equation}

In this work we assume that the stellar mass distribution traces the
luminosity. Thus, we first derive the brightness profile from the mock
image of the central galaxy using MGEs, and use this as the
distribution of the stellar mass. Only the amplitude of the stellar
mass distribution is allowed to vary during the modeling of the
kinematical data, i.e., a constant $M^*/L$ is assumed at all
radii. \citet{Newman2015} conclude that the assumption
  of a constant $M^*/L$ is the main systematic uncertainty in the
  estimation of $\bar{\gamma}_{\rm dm}$. We will discuss the validity
of this assumption in Section~\ref{sec:masslight}.

\begin{figure}
 \includegraphics[width=0.5\textwidth]{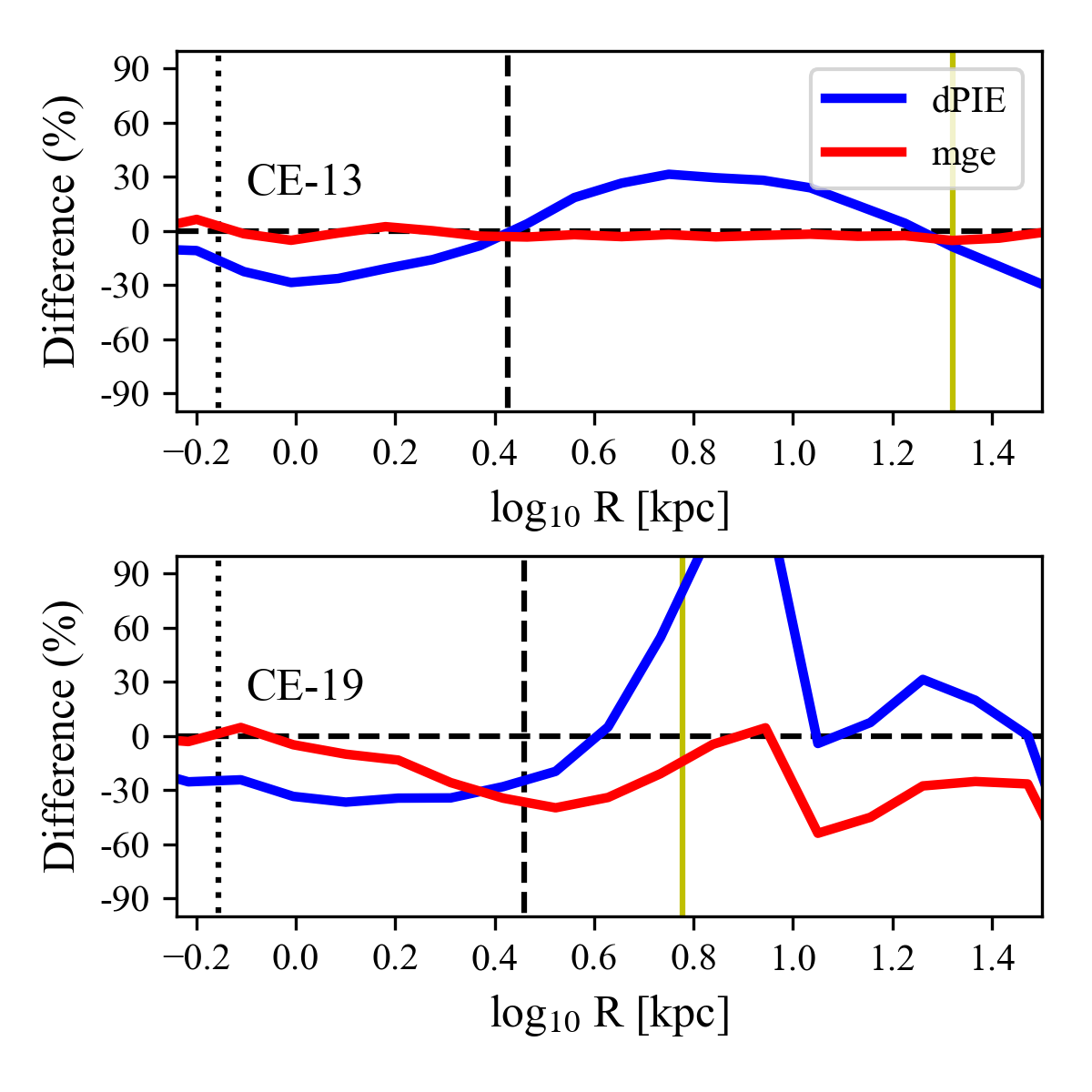}
 \caption{Difference between recovered and true surface luminosity
   profiles for CE-13 and CE-19. Blue lines represent the difference
   between the dPIE and the true profiles. Red lines show the
   difference between the MGE and the true profiles. The vertical yellow lines indicate the outermost radius at which kinematical data are available, which is 21~kpc unless affected
     by satellites along the line-of-sight. The vertical dashed lines
   mark the \citealt{Power2003} radius and the vertical dotted lines
   the softening length.}
 \label{fig:pholum}
 \end{figure}

We compare the quality of the fits to stellar photometry for the dPIE
and MGE models in Fig.~\ref{fig:pholum} for two clusters; the panel
for CE-13 illustrates a typical fit while the panel for CE-19 is the
worst fit amongst 17 clusters.  Clearly, MGE provides a much better
fit than dPIE because it has more free parameters. MGE fits most
clusters within an error of 10\%, while dPIE fits most cluters within
an error of 40\%, which is higher than the errors estimated by
\citet{Newman2013a} ($\sim$ 5\%) from fitting the surface brightness
profiles of their clusters. This discrepancy could be
  due, in part, to differences in the properties of the real and
  simulated galaxies but, as we will show later, it has no affect on
  the inference of the parameters of interest here. Both MGE and dPIE
give a bad fit to CE-19 due to contamination from two line-of-sight
satellites that are very close to the BCG (within 15 kpc).
For JAM we also assume that the dark matter halo follows a gNFW
 profile and the density distribution of the gNFW dark matter halo is
 also expressed as an MGE in the JAM routines.

By requiring that the predicted $v_{\rm rms}$ should be a good match
to the mock galaxy's $v_{\rm rms}$, we can estimate the following six
parameters:
\begin{enumerate}
 \item{the inclination angle, $i$, between the line of sight and the axis of symmetry};
 \item the anisotropy parameter, $\beta_z$, in Equation~(\ref{eq:beta});
 \item the stellar mass-to-light ratio, $M^*/L$;
 \item the three parameters of the dark matter halo density profile:
   $\rho_{\rm s}$, $r_{s}$ and $\gamma$.
\end{enumerate}

\subsection{Model inference with the MCMC method}
\label{sec:MCMC}
According to Bayes' theorem, the posterior likelihood for a set of
parameters, \textbf{p}, given a set of data, \textbf{d}, is: 
 \begin{equation}\label{equ:bayes}
 \text{P}(\textbf{p}|\textbf{d}) = \frac{\text{P}(\textbf{d}|\textbf{p})\text{P}(\textbf{p})}{\text{P}(\textbf{d})}\:,
 \end{equation}
 where $\text{P}(\textbf{d}|\textbf{p})$ is the likelihood and
 P(\textbf{p}) is the prior distribution of the parameters. Combining
 the ``observational'' data together with the models described above,
 we explore the posterior distribution of the model
   parameters using the Markov Chain Monte Carlo (MCMC)
   technique\footnote{We use the ``\textsc{emcee}'' code to
     implement MCMC \citep{Foreman2013}.}.  Assuming the errors are
 independent and Gaussian, the likelihood of a set of parameters is
 proportional to e$^{-\chi^2/2}$, with $\chi^2$ defined as:
\begin{equation}\label{equ:chi}
\chi^2 = \chi_{K}^2 + \chi_{SL}^2 + \chi^{2}_{WL}, 
\end{equation}
where the constraints from kinematics, strong and weak lensing are
described by $\chi^2_K$, $\chi_{SL}^2$ and $\chi^{2}_{WL}$
respectively. Here, $\chi^2_{SL}$ and $\chi^2_{WL}$ take the form
\begin{equation}\label{equ:sl}
\chi_{SL}^2 = \left(\frac{\Sigma(<R_{E})-\Sigma^{'}(<R_{E})}{\sigma_{SL}}\right)^2\:,
\end{equation}
and
\begin{equation}\label{equ:wl}
\chi_{WL}^2 = \sum_{i}\left(\frac{\Delta\Sigma(R)-\Delta\Sigma^{'}(R)}{\sigma_{wl}(R)}\right)^2,
\end{equation}
respectively, where the sum is over 10 data bins. 
$\Sigma(<R_{E})$ is the total enclosed surface mass density, including the baryonic , dark matter and gas components, within the Einstein radius,
and $\Delta\Sigma(R)$ is defined in
Eq.~(\ref{equ:weak}); $\sigma_{SL}=0.05\Sigma^{'}(<R_{E})$ and $\sigma_{wl}=0.4\Delta\Sigma^{'}(R)$ are the 
corresponding errors.  
$\chi^2_{K}$ takes the form
\begin{equation}\label{equ:kinematic}
\chi^2_K = \sum_{i}\left(\frac{v^i_{\rm rms}-v^{'i}_{\rm rms}}{\sigma^i_{\rm rms}}\right)^2\:,
\end{equation}
where the $v^i_{\rm rms}$ is derived through JAM, 
$\sigma^i_{\rm rms}$ is the error and the sum is over 7 data bins. Note that for the sJ model,
$\chi^2_K$ is calculated by substituting the rms velocity,
$v_{\rm rms}$, with the line-of-sight velocity dispersion,
$\sigma_{\rm los}$.

Throughout this paper, we use primed and unprimed quantities to refer to
quantities derived from recovered models and from the original C-EAGLE data,
respectively. Priors for the parameters are listed in Table~(\ref{tab:prior}). We
use uniform priors over reasonable intervals for all parameters,
which are similar to those adopted by
  \citet{Newman2013a}. Note that in this work the ``best-fit''
parameters are given by the median values of the posterior
distributions.

\begin{table}
	\centering
	\caption{Parameter priors. Here, U[a,b] denotes a uniform
          distribution over the interval [a,b] and $\theta$ is the upper
          boundary for $cos(i$) determined from the MGE model.}
         \label{tab:prior} 
	\begin{tabular}{ccc}
		\hline
		\hline
		Parameter & Prior & Unit \\
		cos i & U[0, $\theta$] & \\
		$\beta$ & U[-0.4, 0.4] & \\
		log$_{10}$ $\rho_s$ & U[3, 10] & M$_{\odot}$ \\
		log$_{10}$ r$_s$ & U[${\rm log_{10}(50)}$, 3] & kpc \\
		$\gamma_{\rm gNFW}$ & U[-1.5, 0] & \\
		\hline
		\hline
	\end{tabular}
\end{table}

\section{Results}
\label{sec:results}


\subsection{Recovered density slopes}

\begin{figure*}
	\includegraphics[width=0.85\textwidth]{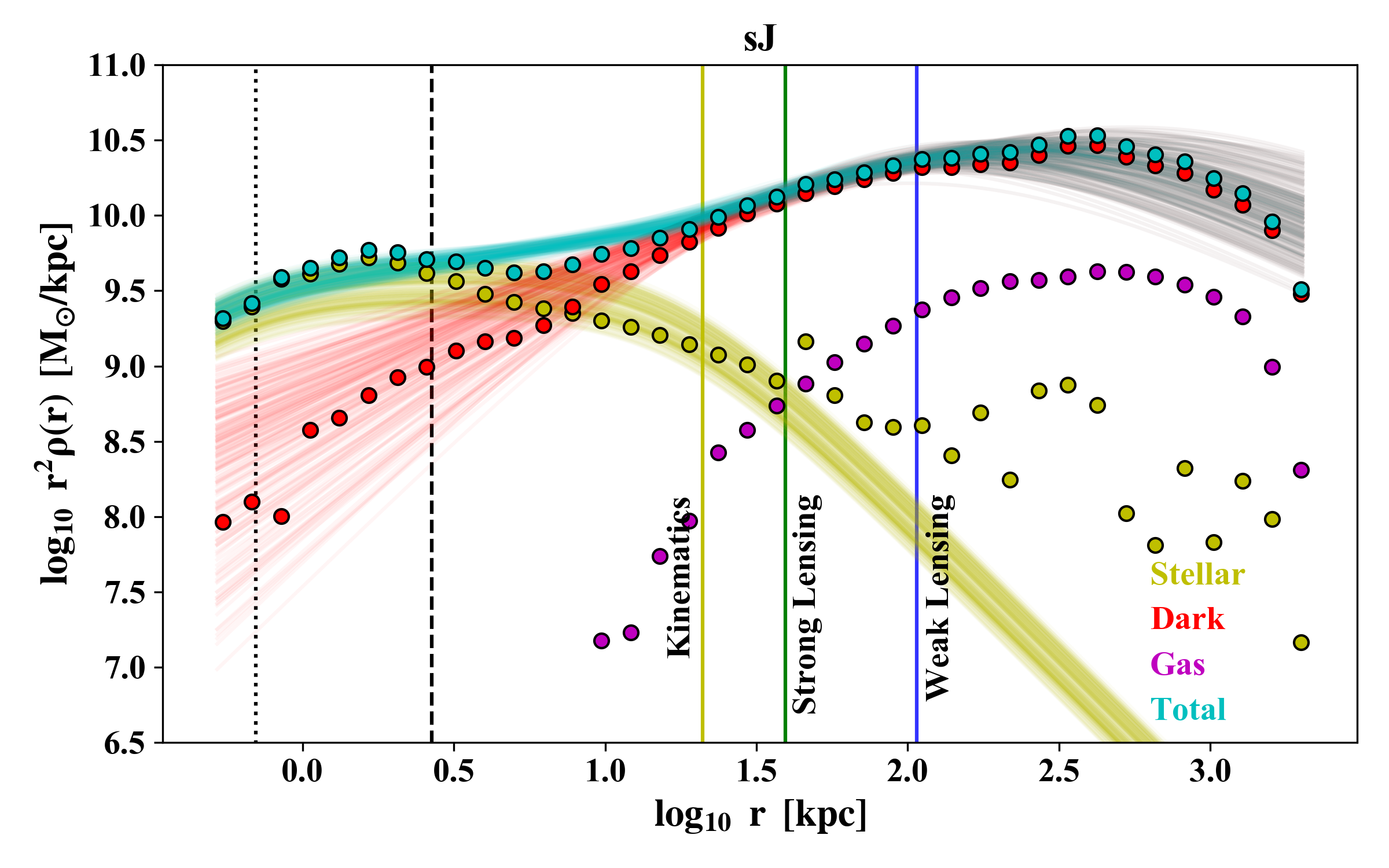}
	\includegraphics[width=0.85\textwidth]{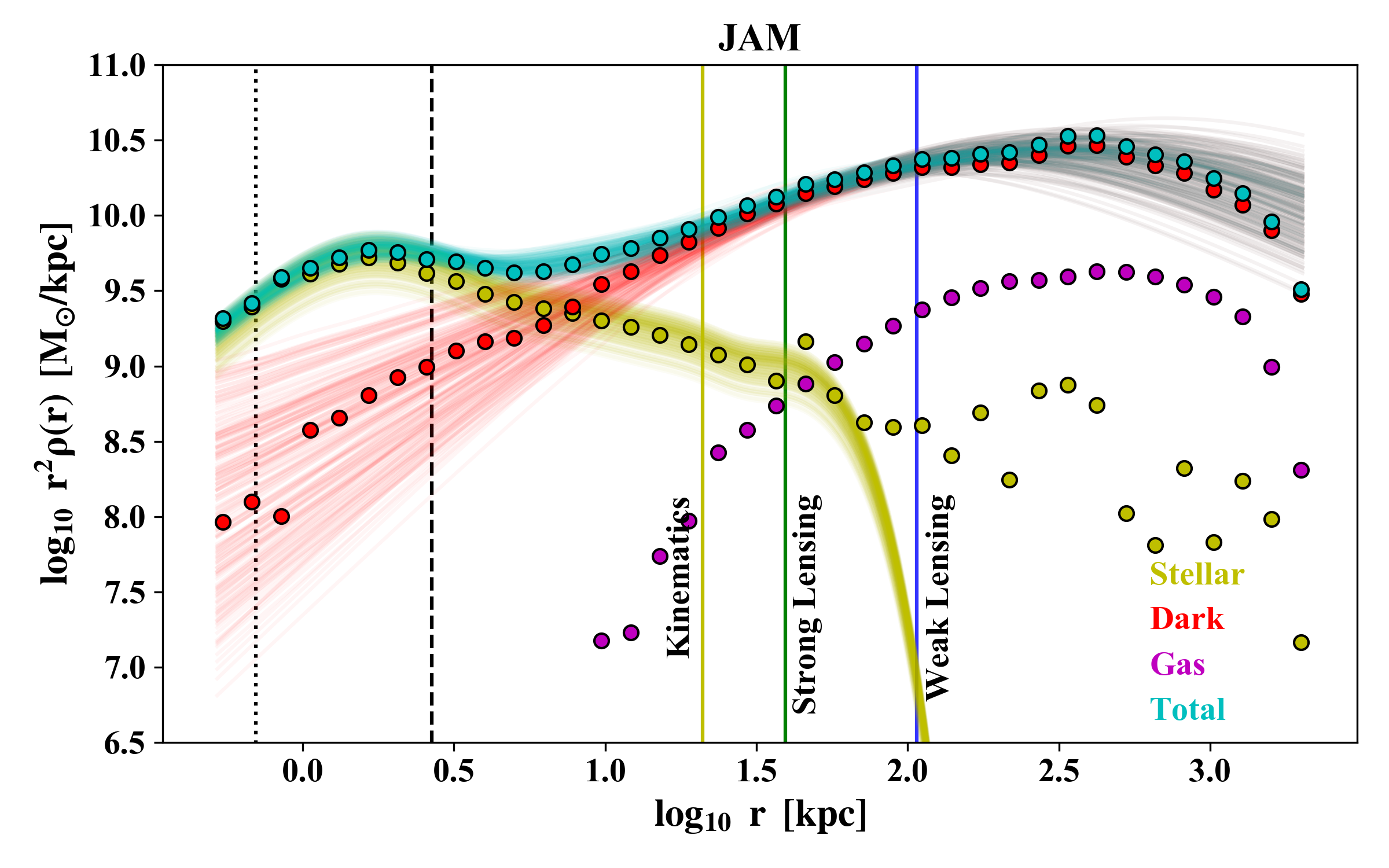}
	\caption{Reconstructed density profiles (r$^{2}\rho$) for
          halo~CE-13. The upper and lower panels show the
          reconstructed profiles using the sJ and JAM models
          respectively. The points show the true density profiles. The
          solid lines show 200 randomly selected reconstructed
          profiles from our MCMC samples. The dark matter, stars, gas
          and total density profiles are plotted in red, yellow,
          magenta and blue respectively. The vertical yellow line
          (r$=21$~kpc) marks the upper bound of the dynamical
          data. The vertical green line marks R$_{\rm Ein}$. Weak
          lensing data exist to the right of the vertical blue
          line. The vertical dashed lines mark the 3D Power {\it et al.}
          radius and the vertical dotted lines the softening length.}
\label{fig:halo13_profiles} 
\end{figure*}

As an example, in Fig.~\ref{fig:halo13_profiles} we compare the
inferred and true density profiles for CE-13. The upper and lower panels show the
results for sJ and JAM respectively. For both models, the recovered
density profiles agree very well with the input ones except for stars
beyond around 100~kpc. Since our fiducial stellar mass model assumes a
constant mass-to-light ratio and our dPIE (MGE) fit to the light
distribution is restricted to 100 kpc, the discrepancy beyond this
radius is to be expected. Note that although the two models give very
similar profiles for CE-13, there are still differences in the inner
dark matter profiles where sJ tends to overestimate the mass of dark
matter.

\begin{figure*}
	\includegraphics[width=0.7\textwidth]{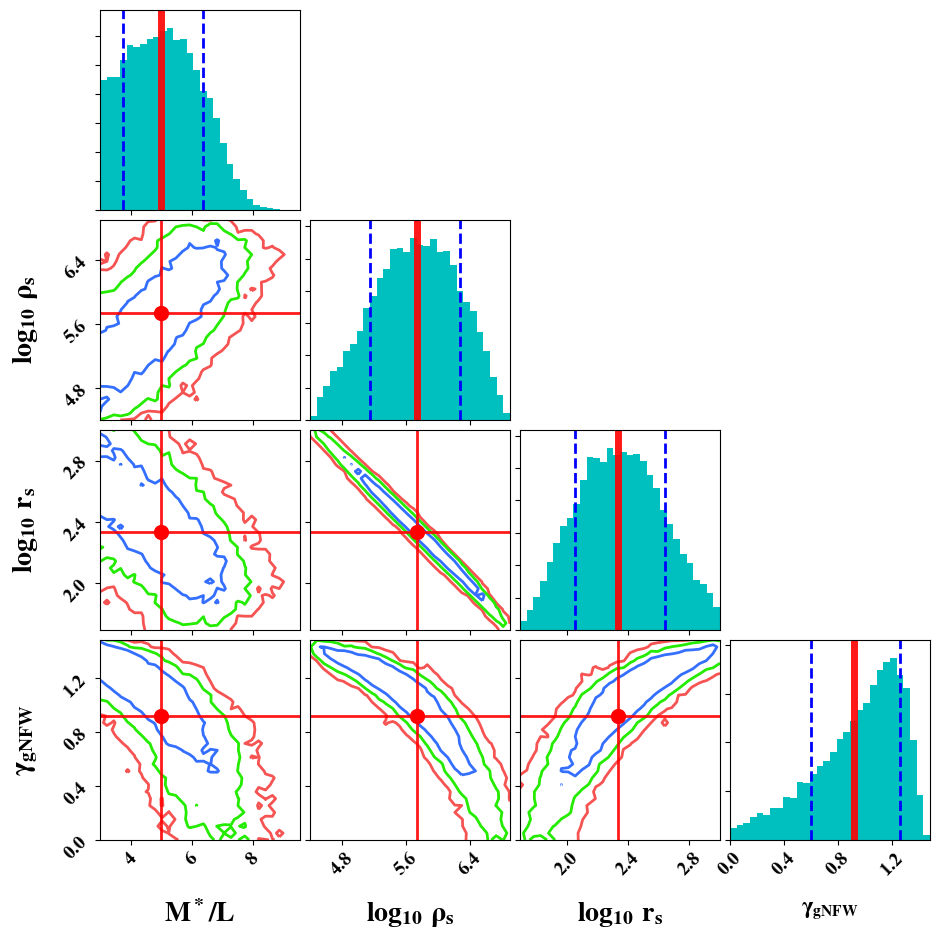}
	\caption{Posterior distributions of model parameters for the
          sJ + lensing analysis. In the panels with contours,
          true values of the parameters are marked with red
          dots. Blue, green and red lines represent 1-, 2- and
          3$\sigma$ regions, respectively. In the marginalized
          distributions the input values are marked with vertical red
          solid lines and the 84\% and 16\% percentiles with vertical
          dashed blue lines.}
\label{fig:mcmc_sJ} 
\end{figure*}

\begin{figure*}
	\includegraphics[width=0.9\textwidth]{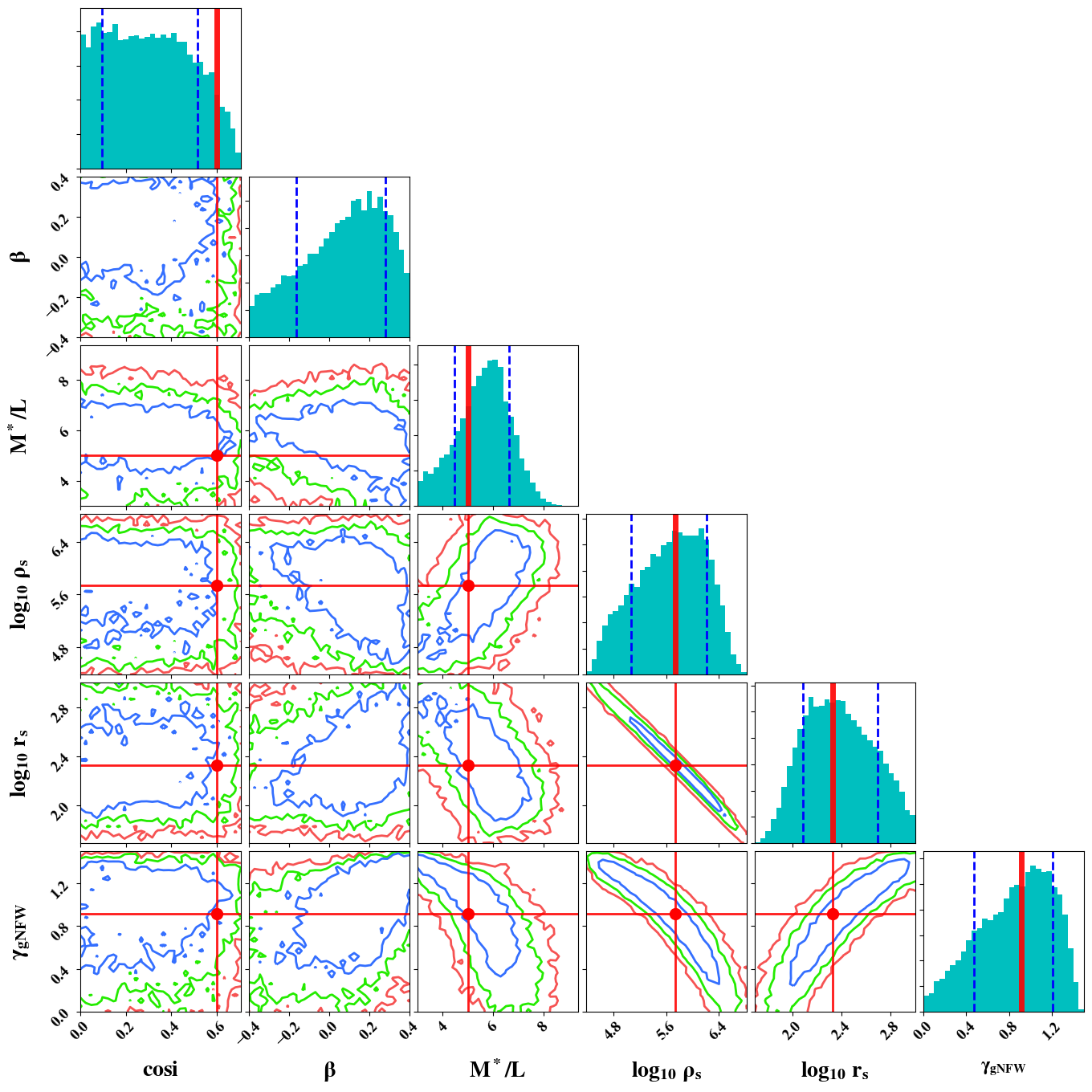}
	\caption{Posterior distributions of model parameters for the
          JAM + lensing analysis. In the panels with contours, the
          true values of the parameters are marked with red
          dots. Blue, green and red lines represent 1-, 2- and
          3$\sigma$ regions, respectively. In the marginalized
          distributions the input values are marked with vertical red
          solid lines and the 84\% and 16\% percentiles with vertical
          dashed blue lines. The true value of $\beta$
            is not shown in the plot because it lies outside the prior
            range; this happens only in the case of CE-13.}
\label{fig:halo13} 
\end{figure*}

In Figs.~\ref{fig:mcmc_sJ} and Fig.~\ref{fig:halo13} we present the
posterior distributions of the model parameters for sJ + lensing
and JAM + lensing analyses, respectively, for CE-13. For both sJ and JAM +
lensing analyses, significant degeneracies among the three parameters of the
gNFW fit can be clearly seen in the contours. To compare
  our best-fit gNFW profiles with the input dark matter profiles, we
  also fit the latter between 1 kpc and R$_{200}$ to get the ``true''
  input values of the gNFW parameters. Different choices for the
  radial range in the fit and the weighting scheme can lead to
  slightly different best-fit values because of degeneracies amongst
  gNFW parameters. For example, the values of $\gamma_{\rm gNFW}$
  inferred from fitting to the mass profiles are systematically
  smaller than those inferred from fitting to the density profiles by
  $\sim$ 0.12. These systematic differences are well below the
  statistical errors of the estimates derived from kinematics +
  lensing analysis we have carried out.

To compare the total inner density slope, we additionally define a
mass-weighted density slope in the same way as \citet{Dutton2014b} and
\citet{Newman2015}:
\begin{equation}\label{equ:rho_bar}
\color{black}
\bar{\gamma}_{\rm dm} \equiv -\frac{1}{M(R_e)}\int_0^{R_e} 4\pi r^2 \rho(r) \frac{d\log{\rho}}{d\log{r}}dr= 3 - \frac{4\pi R_e^3\rho(R_e)}{M(R_e)} \:,
\end{equation}
where $\rho(r)$ and $M(r)$ are the cluster's total density and mass
profiles. Similarly, we define the mass-weighted dark matter density
slope, $\bar{\gamma}_{\rm dm}$, by using the dark matter $\rho(r)$ and
$M(r)$ density profiles in Eqn.~\ref{equ:rho_bar}. In the case where
the dark matter scale radius, $r_s \gg R_{\rm e}$, the dark matter
density slope within $R_{\rm e}$ follows a power-law distribution and
the asymptotic slope $\gamma_{\rm gNFW}$ is equivalent to
$\bar{\gamma}_{\rm dm}$. Note that the ``true"
  mass-weighted slope is calculated directly from the simulation data
  rather than derived from a fit to the profile.

\begin{figure*}
	\includegraphics[width=0.97\textwidth]{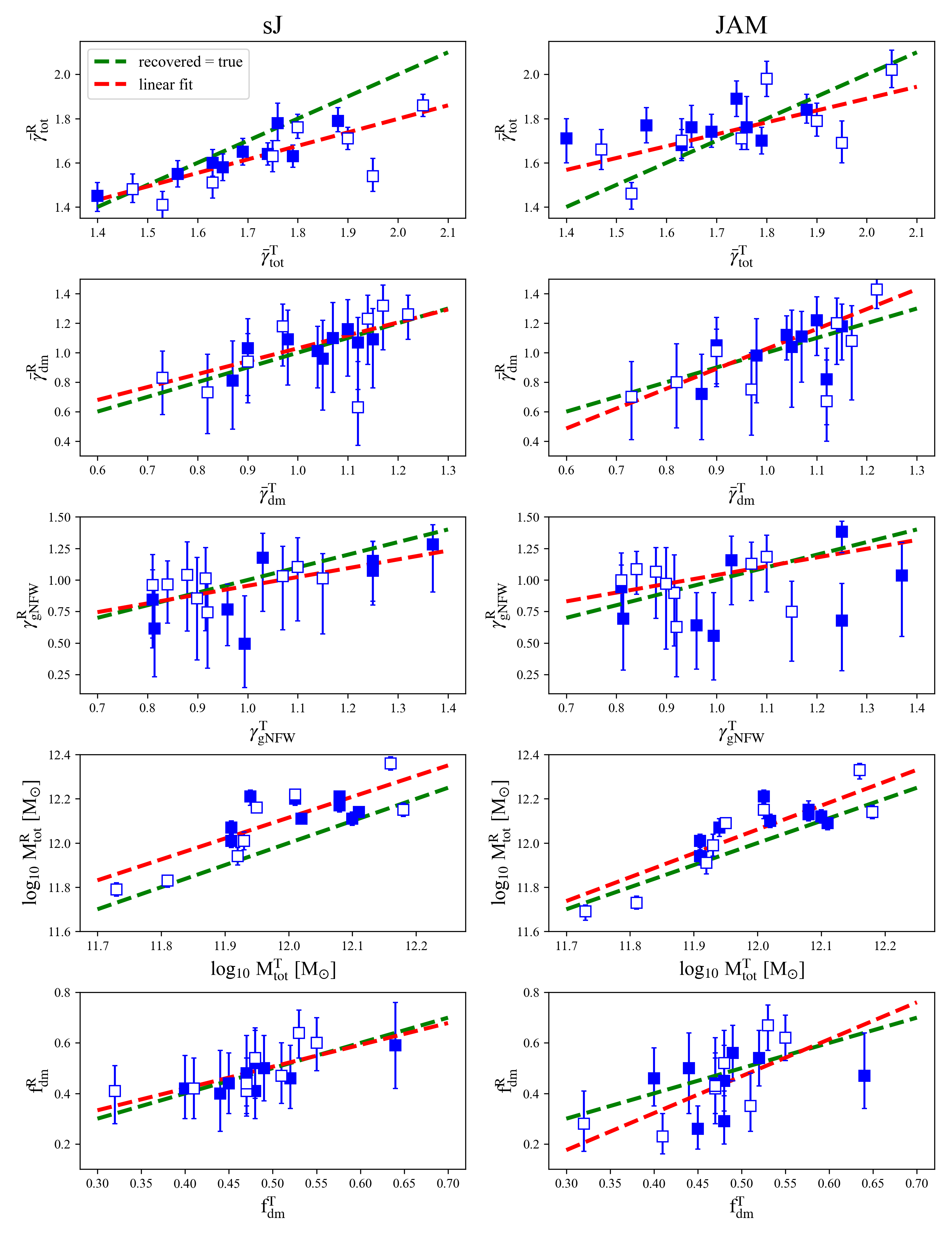}
	\caption{Comparison between the true and best-fit values of
          $\bar{\gamma}_{\rm tot}$, $\bar{\gamma}_{\rm dm}$,
          $\gamma_{\rm gNFW}$, log(M$_{\rm tot}$) and f$_{\rm
            dm}$. The left and right columns show results for sJ and
          JAM respectively. The $x$-axis is the true value and the
          $y$-axis the best-fit value. (The best-fit and true values
          are denoted by superscripts ``R" and ``T" respectively.) The
          solid squares represent relaxed and the empty squares
          unrelaxed clusters. The red dashed lines are the best linear
          fits to the true {\it vs} best-fit values. The green dashed
          lines indicate equality.}
\label{fig:compare}
\end{figure*}

In Fig.~\ref{fig:compare} we compare the true and the best-fit values
of several key parameters: $\gamma_{\rm gNFW}$, the asymptotic density
slope of the dark matter halo; $\bar{\gamma}_{\rm tot}$, the
mass-weighted average density slope within $R_{\rm e}$ for the total
mass distribution; $\bar{\gamma}_{\rm dm}$, the mass-weighted average
density slope within $R_{\rm e}$ for the dark matter distribution;
$M_{\rm tot}$, the total mass within $R_{\rm e}$; $f_{\rm dm}$, the dark matter fraction within
$R_{\rm e}$, for sJ + lensing analysis (left column) and JAM + lensing (right column) respectively. We denote the best-fit and true
values with superscript ``R" and ``T" respectively. 

To illustrate clearly the trend between best-fit and true values, the
green dashed lines indicate equality; the red dashed lines are the
linear relation between best-fit and true values. For
  both models, the mass-averaged dark matter density slopes,
  $\bar{\gamma}_{\rm dm}$, and $\gamma_{\rm gNFW}$ are reasonably well
  constrained. For the total mass within R$_{\rm e}$, M$_{\rm tot}$
is overestimated by 0.1 $\sim$ 0.2 dex for many
clusters. Interestingly, the best-fit total density slope,
$\bar{\gamma}_{\rm tot}$, behaves very differently between the two
models. JAM tends to overestimate the total density slope at small
masses, while sJ systematically underestimates the total density slope
at high masses. For the dark matter fraction both models provide an
unbiased recovery, with sJ showing smaller variance than JAM. The
parameter values in Fig.~6 are also listed in Tables~\ref{tab:results}
and~\ref{tab:results_sJ}.
  
To investigate whether the recovered mass depends on the dynamical
state of the cluster, we classify the C-EAGLE clusters as
relaxed or unrelaxed using the information provided in Table~A2 of
\citet{Barnes2017}. A cluster is defined as relaxed if the kinetic
energy of the gas is less than 10\% of the total thermal energy within
$R_{500}$. In Fig.~\ref{fig:compare} we use filled squares to indicate
relaxed clusters and empty squares to indicate unrelaxed
clusters. Overall, the quality of the recovery is independent of the
dynamical state of the cluster.

%
%
%

%

\subsection{Comparison with observations}

\begin{figure}
  \includegraphics[width=0.5\textwidth]{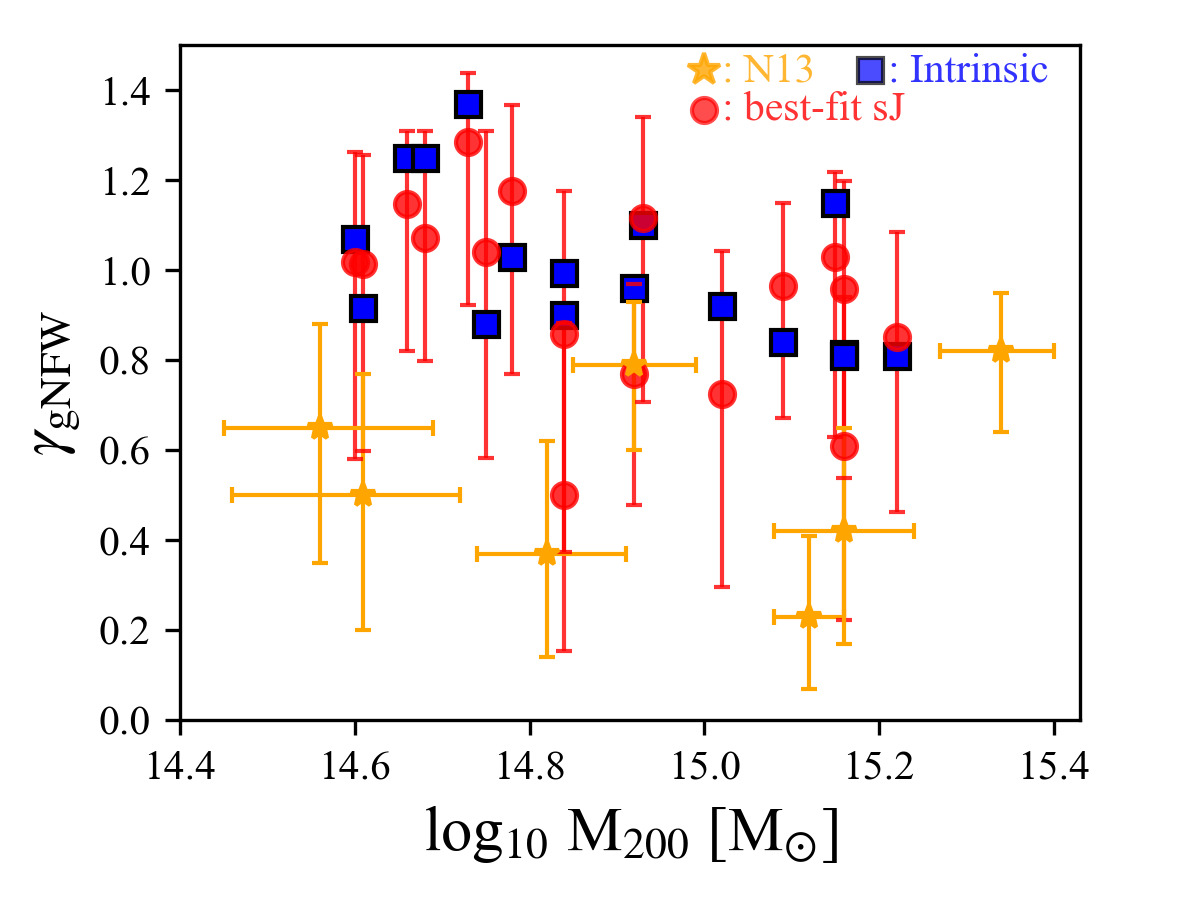}
  \caption{$\gamma_{\rm gNFW}$ as a function of M$_{200}$. Blue
    squares show the true values for the C-EAGLE clusters and the red
    circles the best-fit values from the sJeans + lensing analysis. The
    observational estimates from \citet{Newman2013b} are shown as
    yellow stars.}
\label{fig:gamma_comparison} 
\end{figure}

In this section, we compare our C-EAGLE mocks with the observed
clusters of \citet{Newman2013a,Newman2013b,Newman2015}. Fig.~\ref{fig:gamma_comparison} shows the best-fit asymptotic dark
matter density slopes, $\gamma_{\rm gNFW}$, as a function of the
cluster mass, $M_{\rm 200}$, derived from our mock cluster data and
from the observations of \citet{Newman2013b}. For comparison, we also plot the input
values of $\gamma_{\rm gNFW}$, which we derived by fitting the gNFW
profile directly to the simulation data.

The true asymptotic dark matter density slopes of the C-EAGLE clusters
have values $\sim 1$ at $10^{14.5}$ M$_{\odot}$ and decrease slowly to
$\sim 0.8$ at $10^{15}$ M$_{\odot}$. These are significantly higher
than the observational results of \citet{Newman2013b}, for which the
mean value is $0.50\pm 0.13$ (with an estimated systematic error of
0.14). For both the sJ and JAM + lensing analyses the recovered values
of $\gamma_{\rm gNFW}$ agree well with the input ones, and both are
systematically higher than those inferred from the observational
data. To be specific, we use bootstrap methods to choose 7
(the same number of clusters as in \citet{Newman2013a})
asymptotic slopes, $\gamma_{\rm gNFW}$, randomly from the posterior
distributions of $\gamma_{\rm gNFW}$ for all 17
clusters to derive the joint constraint on the mean value of $\gamma_{\rm gNFW}$. The method we use is different from the
  method used by \citet{Newman2013b} who multiplied the posterior
  distributions of $\gamma_{\rm gNFW}$ together, implicitly assuming
  that these distributions are the same for all clusters; this is not
  necessarily the case and the product can be strongly affected by
  inclusion of one or two clusters with a very different
  $\gamma_{\rm gNFW}$ distribution.  In fact, \citet{Newman2013b}
  point out that excluding the cluster with the lowest
  $\gamma_{\rm gNFW}$ (A2537), the mean $\gamma_{\rm gNFW}$ would
  change by $\sim$ 40\% from 0.50 to 0.69. Using the bootstrap method,
  we find probabilities of 3.5\%, 17.1\% and 49.1\% for the mean value
  of those randomly chosen 7 asymptotic slopes to lie within the
  1$\sigma$ ($0.50\pm0.13$), 2$\sigma$ ($0.50\pm0.26$) and 3$\sigma$
  ($0.50\pm0.39$) ranges of the \citet{Newman2013b} results,
  respectively. In this comparison, we combine the constraints from
the kinematic, strong lensing and weak lensing data as was done by
\citet{Newman2013b} and, like them, we use the sJ model for the
dynamical analysis. We assume similar uncertainties for the kinematics
and strong lensing as in the observational study and reasonable
uncertainties for the weak lensing constraint as shown by
  Fig.5 in \citet{Newman2013a}. Thus, the discrepancy between the
observed inner density slopes and those of the C-EAGLE clusters is
unlikely to be due entirely to systematics in the method itself.

\begin{figure}
    \includegraphics[width=0.5\textwidth]{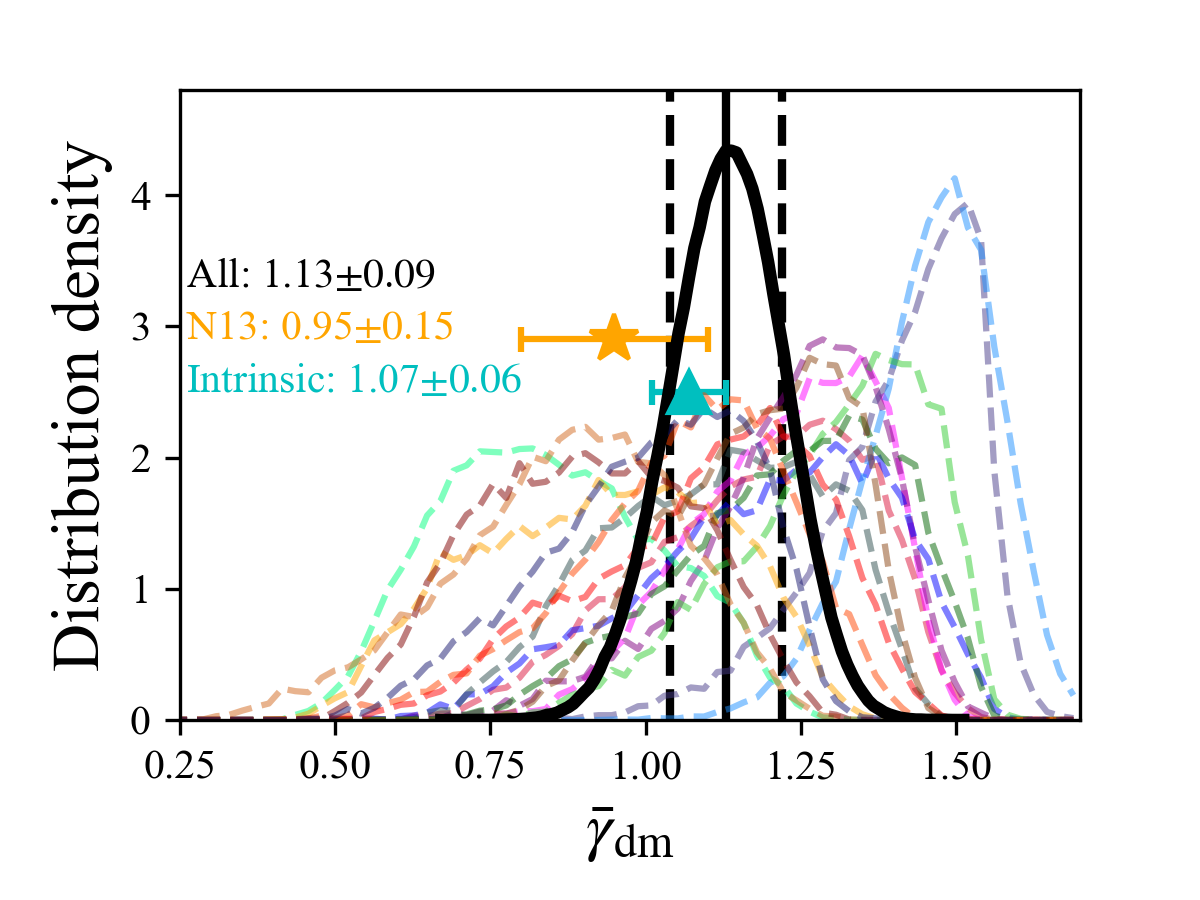}
    \caption{Marginalized posterior distributions of
      $\bar{\gamma}_{\rm dm}$ (at 44 kpc) obtained from the sJ + lensing analysis
      for each C-EAGLE cluster (dashed lines). The vertical solid black
      line shows the mean value of $\bar{\gamma}_{\rm dm}$. The solid
      curve shows the joint constraint on the mean value of
      $\bar{\gamma}_{\rm dm}$ and the vertical dashed lines the
      16\% and 84\% percentiles. The yellow star with an error bar is
      the corresponding result of \citet{Newman2013b}. The cyan
      triangle is the true value for our C-EAGLE sample, with the bar
      spanning the error in the mean. The values given in the legend are
      the most probable values for the mean value of
      $\bar{\gamma}_{\rm dm}$.}
\label{fig:ngMgamma}
\end{figure}

Interestingly, the simulation and observational results
  agree well if we compare the mean values of the mass-weighted
  mean density slopes within the effective radius,
  $\bar{\gamma}_{\rm dm}$, instead of the asymptotic
  $\gamma_{\rm gNFW}$. Since the effective radii of the central
  galaxies of the C-EAGLE clusters are smaller than those of the
  \citet{Newman2013a} sample, roughly 44 kpc, to be consistent we
  measure $\bar{\gamma}_{\rm dm}$ for our clusters at 44 kpc using
  Eq.~\ref{equ:rho_bar}. (If not explicitly stated,
  $\bar{\gamma}_{\rm dm}$ is taken to mean the value at the effective
  radius of the C-EAGLE cluster.) In Fig.~\ref{fig:ngMgamma} we show
  (with dashed lines) the posterior distribution of
  $\bar{\gamma}_{\rm dm}$ derived by the sJ + lensing analysis for
  each C-EAGLE cluster. We also mark with a vertical solid black line
  the mean value of $\bar{\gamma}_{\rm dm}$. To explore the spread in
  the mean, we again use a bootstrap method to draw 7 values randomly
  from the posterior distribution of $\bar{\gamma}_{\rm dm}$. The
  solid black line in the figure shows the distribution from the
  bootstrap and the vertical dashed black lines its 16\% and 84\%
  percentiles. We find a mean $\bar{\gamma}_{\rm dm} =
  1.13\pm0.09$. The mean and error of the true values of
  $\bar{\gamma}_{\rm dm}$ for the C-EAGLE clusters are shown as a cyan
  triangle and error bar. The yellow star and error bar show
  the mass-weighted mean dark matter slope for the sample of
  \citet{Newman2013a, Newman2013b}, taken directly from Fig.~15 of
  \citet{Newman2015}. It differs from with the true and estimated mean values for the 
  C-EAGLE clusters by less than $\sim 1\sigma$.

\begin{figure}
\includegraphics[width=0.5\textwidth]{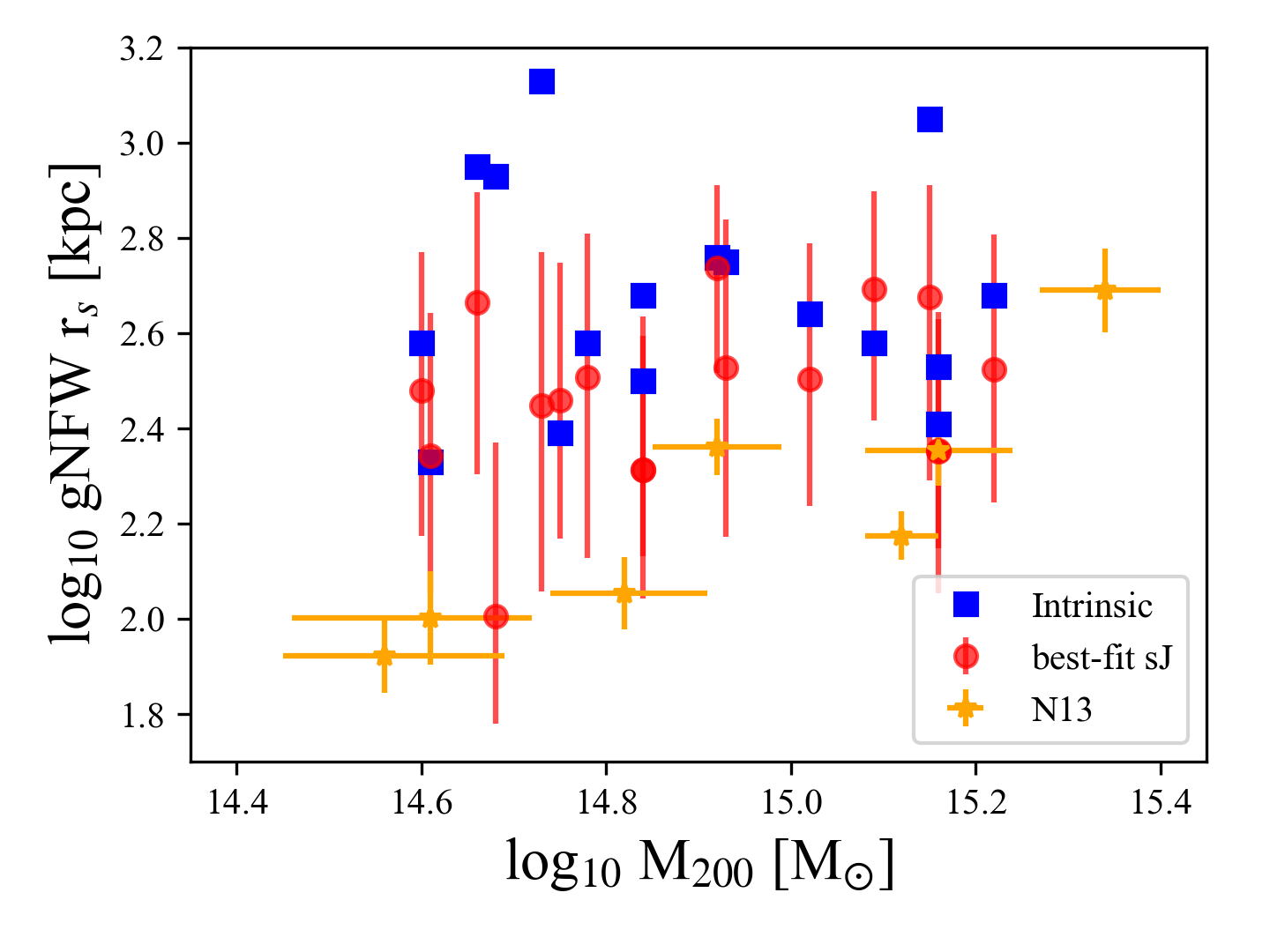}
\includegraphics[width=0.5\textwidth]{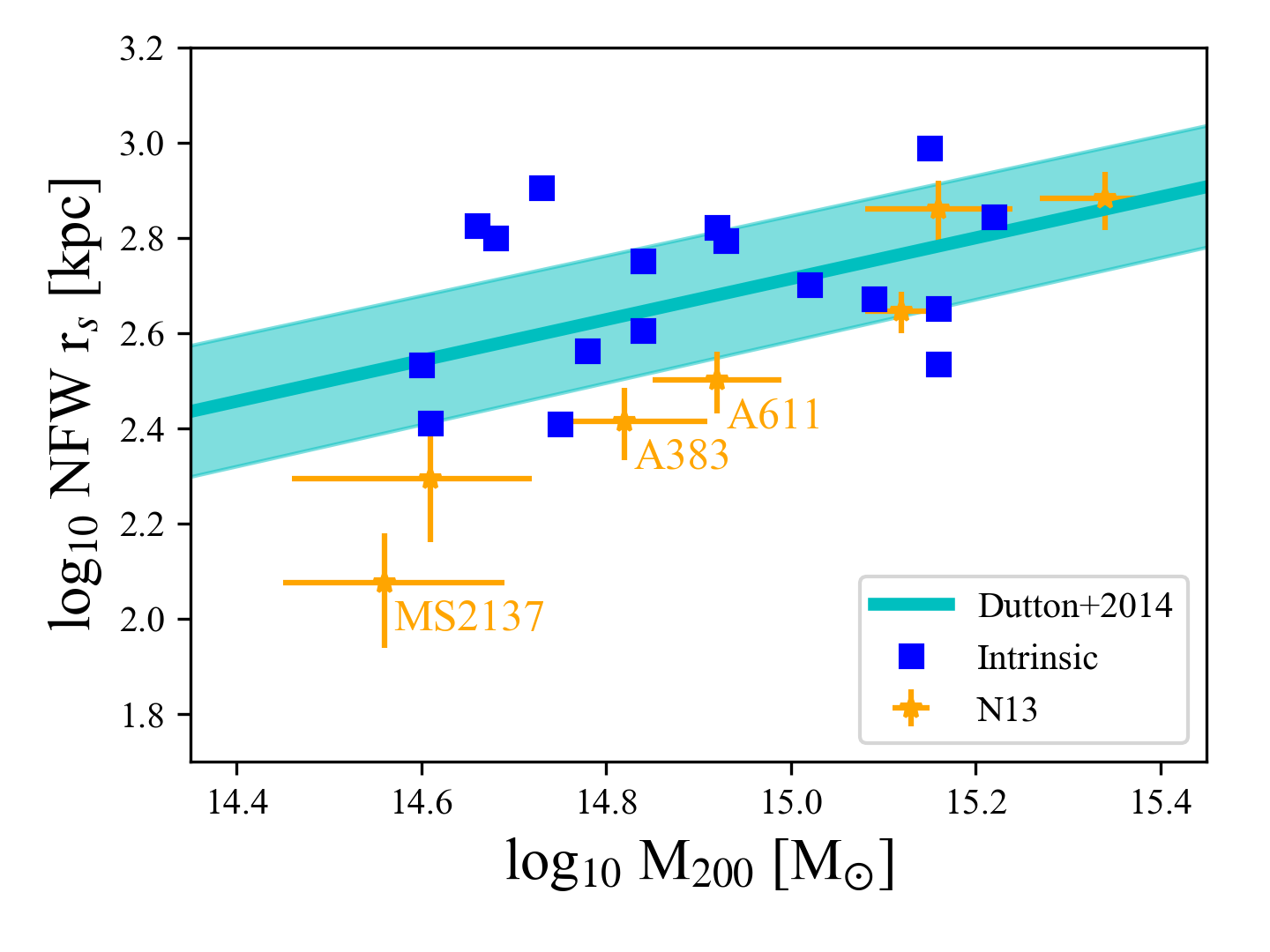}
\caption{\textbf{Upper panel:} gNFW scale radius as a function of
  M$_{200}$. The blue squares are the true values of the gNFW $r_s$
  for the C-EAGLE clusters; the yellow stars with error bars are the
  gNFW $r_s$ values of the \citet{Newman2013b} clusters (estimated
  from the Appendix of \citealt{Newman2013a}); and the red circles are
  the best-fit gNFW $r_s$ values for the C-EAGLE clusters from the sJ
  + lensing analysis. \textbf{Lower panel:} NFW scale radius as a
  function of M$_{200}$. The blue squares are the true values for the
  C-EAGLE clusters; the yellow stars are the values for the
  \citet{Newman2013a} clusters (from their Table~8). The cyan line is
  the mass-concentration relation for relaxed haloes at redshift 0.2
  \citep{Dutton2014} , with the shaded region showing the
  corresponding 1-$\sigma$ scatter \citep{Neto2007}.}
\label{fig:rs_plot} 
\end{figure}

Why are the observed values of $\gamma_{\rm gNFW}$ much smaller than
those from the C-EAGLE simulations, while the respective values of
$\bar{\gamma}_{\rm dm}$ agree? As we discussed before, the
mass-weighted density slope,
$\bar{\gamma}_{\rm dm} \simeq \gamma_{\rm gNFW}$ when
$r_s \gg R_{\rm e}$. Thus, the significant difference between the two
measures of slope in the data of \citet{Newman2013a, Newman2013b}
implies that the observed clusters have smaller inferred values of
$r_s$ than the C-EAGLE clusters. To confirm this point, in the
  upper panel of Fig.~\ref{fig:rs_plot}, we plot the values of the
  gNFW scale radius as a function of M$_{200}$. For the sJ + lensing
  analysis, the best-fit values of $r_s$ agree well with the true
  values. But, as we can see, 4 out of the 7 clusters in
  \citet{Newman2013a, Newman2013b} have inferred values of the gNFW
  $r_s$ smaller than the smallest intrinsic value in the C-EAGLE
  sample, which is around 220~kpc. A simple interpretation of this
  discrepancy is that it reflects differences in the density
  distribution in the outer parts of the simulated and real clusters.
  However, since there is a degeneracy between $r_s$ and
  $\gamma_{\rm gNFW}$, the small inferred values of the gNFW $r_s$
  could reflect the presence of a central core in the cluster dark
  matter distribution.  To resolve this ambiguity we compare the
  estimated values of the {\it NFW} $r_s$ (from Table~8 in
  \citealt{Newman2013a}) with those found in cluster simulations (blue
  squares for C-EAGLE and the cyan line from
  \citealt{Dutton2014}). The observational estimates of the NFW $r_s$
  are obtained exclusively from the strong and weak lensing data which
  pertain to regions far from the centre. As we can see, the estimated
  values of the NFW $r_s$ for the three most massive clusters in the
  Newman et al. sample agree well with the simulations, but those for
  the three smaller clusters in the sample are significantly smaller
  than in any of the simulated clusters, pointing to real differences
  in the outer density profiles of the real and simulated clusters.

\subsection{Importance of lensing constraints}

It is worth pointing out that the lensing data plays a crucial role in
constraining the mass model. Although these data probe only the outer
parts of the density profile, the strong degeneracy
  amongst the three parameters of the gNFW profile implies that they
  are also important for constraining the inner parts of the profile,
  and help improve the precision of the decomposition of the stellar
  and dark mass components. Poor or biased lensing measurements may
lead to the inference of incorrect dark matter slopes.

\subsubsection{Tests with kinematics alone}
\begin{figure}
    \includegraphics[width=0.5\textwidth]{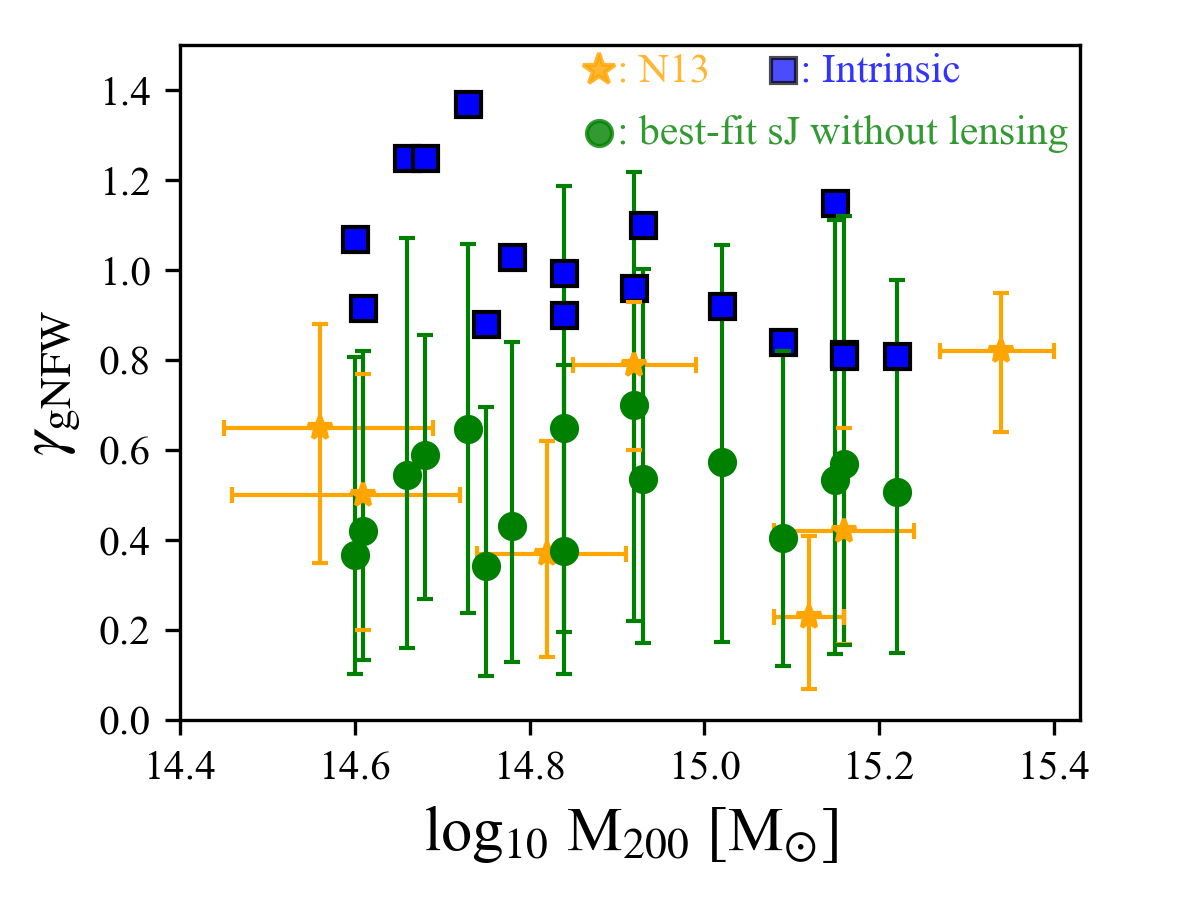}
    \caption{Comparison between the best-fit values of
      $\gamma_{\rm gNFW}$ for the C-EAGLE clusters from the sJ
      analysis alone (ignoring lensing data; green circles), the
      observational estimates from \citet{Newman2013b} (orange stars)
      and the fiducial values of C-EAGLE clusters (blue squares).}
\label{fig:gamma_M200_nl}
\end{figure}

In Fig.~\ref{fig:gamma_M200_nl} we show the best-fit
$\gamma_{\rm gNFW}$ values for the C-EAGLE clusters derived from 
kinematical data alone. For the sJ model, the median value of the
best-fit asymptotic slope, $\gamma_{\rm gNFW}$, is 0.54, which is
significantly smaller than the true value. The JAM model produces a
slightly more accurate result, $\gamma_{\rm gNFW}=0.61$, but this
still significantly underestimates the true density slopes.

\begin{figure}
	\includegraphics[width=0.5\textwidth]{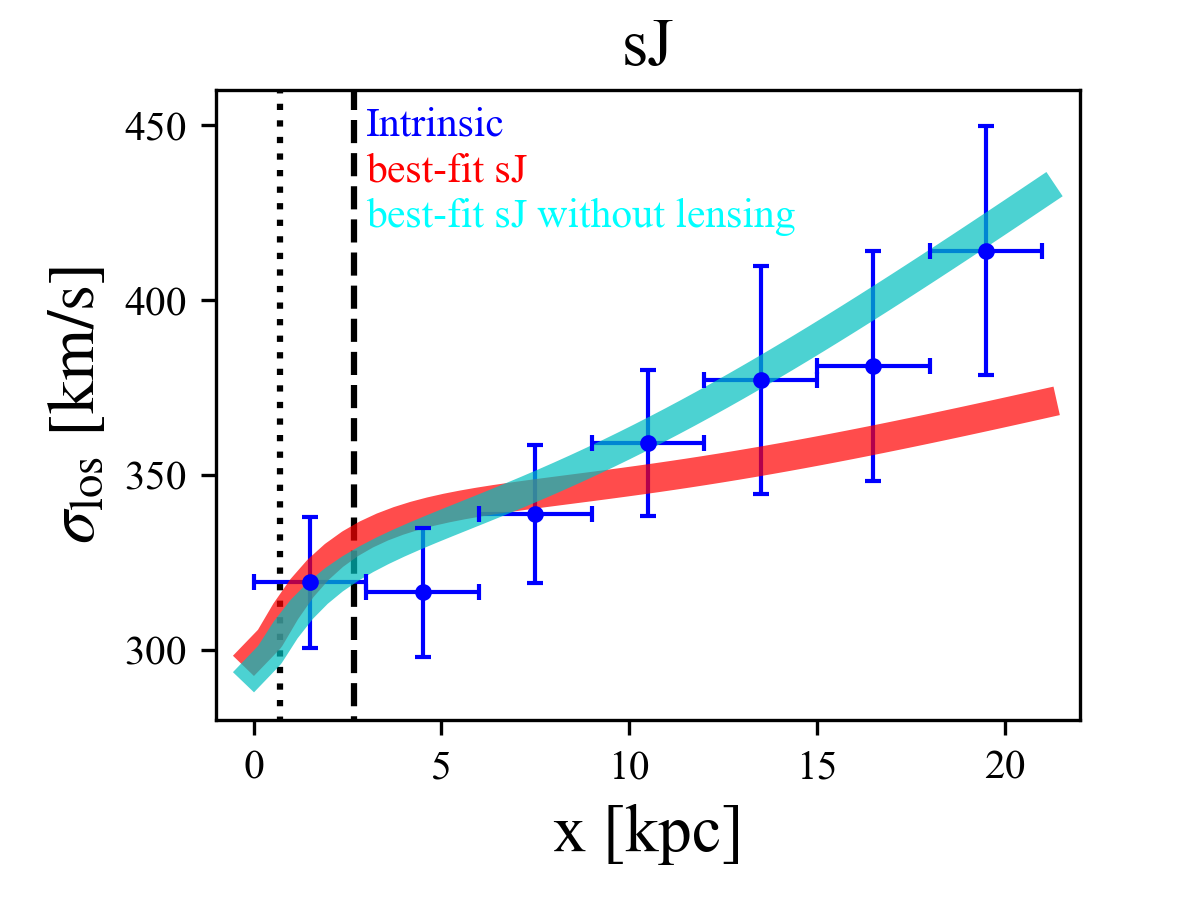}
	\includegraphics[width=0.5\textwidth]{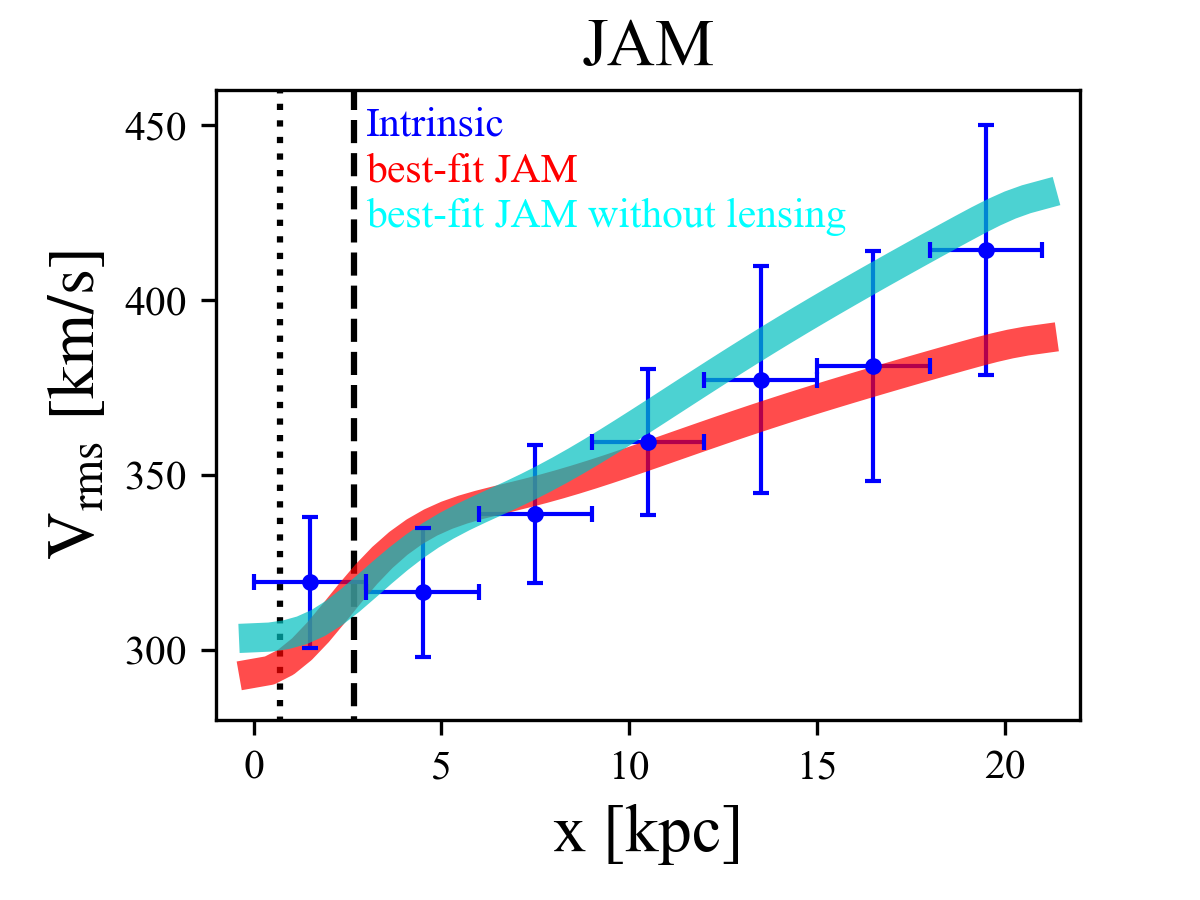}
	\caption{Comparison between true and recovered dynamical
          quantities for both the sJ and JAM models for
            CE-13. The upper panel shows the velocity dispersion of
          the stars and the bottom panel the rms velocity along the
          line-of-sight.  The blue circles with error bars show the
          true values; the red lines represent the dispersions
          inferred from the full model and cyan lines those ignoring
          the lensing constraints. The vertical dotted lines mark the
          softening length and the vertical dashed lines the 3D Power
          {\it et al.} radius. The width of the error
            bars represents the size of the bins used to derive the
            input kinematics.}
\label{fig:rms_plot} 
\end{figure}

Why does dynamical modeling alone fail to reproduce the input
$\gamma_{\rm gNFW}$? The reason may lie in the lack of information
about the halo profile contained by the dynamical data, which are
restricted to the inner halo. In Fig~\ref{fig:rms_plot}, we show dynamical quantities for CE-13 inferred from the sJ and JAM
models. The blue points with error bars are the true values and the
red lines are best-fit results including the lensing constraints. Both
sJ and JAM fit most of the dynamical data within the errors, with JAM
providing a better fit than sJ. Both models, however, underestimate
the velocity dispersion at $x\sim 20$~kpc (even though they both
accurately recover the true total density profiles).  Ignoring the
lensing constraints (cyan lines), both models overestimate the
velocity dispersions at large radii. This is because, confined to the
central parts of the cluster, the dynamical data alone, without the
lensing data, cannot constrain the gNFW profile,
  especially the value of $r_s$. The MCMC fitting then tends to
zero-in onto a shallower dark matter density profile slope than the
true value, which increases the velocity dispersion in the outer
regions, where it is underestimated by the full model. This explains
the bias in $\gamma_{\rm gNFW}$ seen in Fig.~\ref{fig:gamma_M200_nl}.

\subsubsection{Tests with biased weak lensing}

In the last section we showed that the lensing
  measurements serve to anchor the constraints on the total density
  profile. Biased lensing measurements are therefore likely to lead to
  biased estimates of $\gamma_{\rm gNFW}$.

  Interestingly, recent studies using weak lensing data of high
  quality find larger values of the NFW scale radius, $r_s$, for some
  of the clusters included in the sample of \citet{Newman2013a}.  In
  table~\ref{tab:NFW_rs}, we compare lensing measurements of $r_s$,
  obtained from NFW fits, for three clusters by \citet{Merten2015,
    Umetsu2016} with the results of \citet{Newman2013a} (see Table~8
  in \citet{Newman2013a}, Table~6 in \citet{Merten2015} and Table~2 in
  \citet{Umetsu2016})\footnote{For consistency, we only compare
    parameters for spherical NFW haloes.}. For these three clusters,
  the values of the scale radii measured by \citet{Newman2013a} are
  smaller than the more recent measurements by the other authors by
  30\% to $\sim$ 700\%. For a comparison with simulation
    predictions, we have also marked those three clusters in the lower
    panel of Fig.~\ref{fig:rs_plot}.

  To explore how the best-fit values of $\gamma_{\rm gNFW}$ are
  affected when the lensing measurements return a profile with too
  small a value of $r_s$, we perform the following test. We first
  obtain best-fit NFW profiles using unbiased weak lensing
  ``measurements'' of the C-EAGLE clusters. Next, without changing the
  value of M$_{200}$, we decrease the scale radius of the best-fit NFW
  profile by 50\%, which is approximately the average difference
  between the results of \citet{Newman2013a} and those of
  \citet{Merten2015} and \citet{Umetsu2016}. We then generate weak
  lensing measurements using these artificially biased NFW profiles
  with the same error bars as the fiducial ones. Finally, we combine
  the fiducial stellar kinematical data and strong lensing data with the artificially biased weak lensing data to constrain the mass models.

\begin{figure}
 	\includegraphics[width=0.5\textwidth]{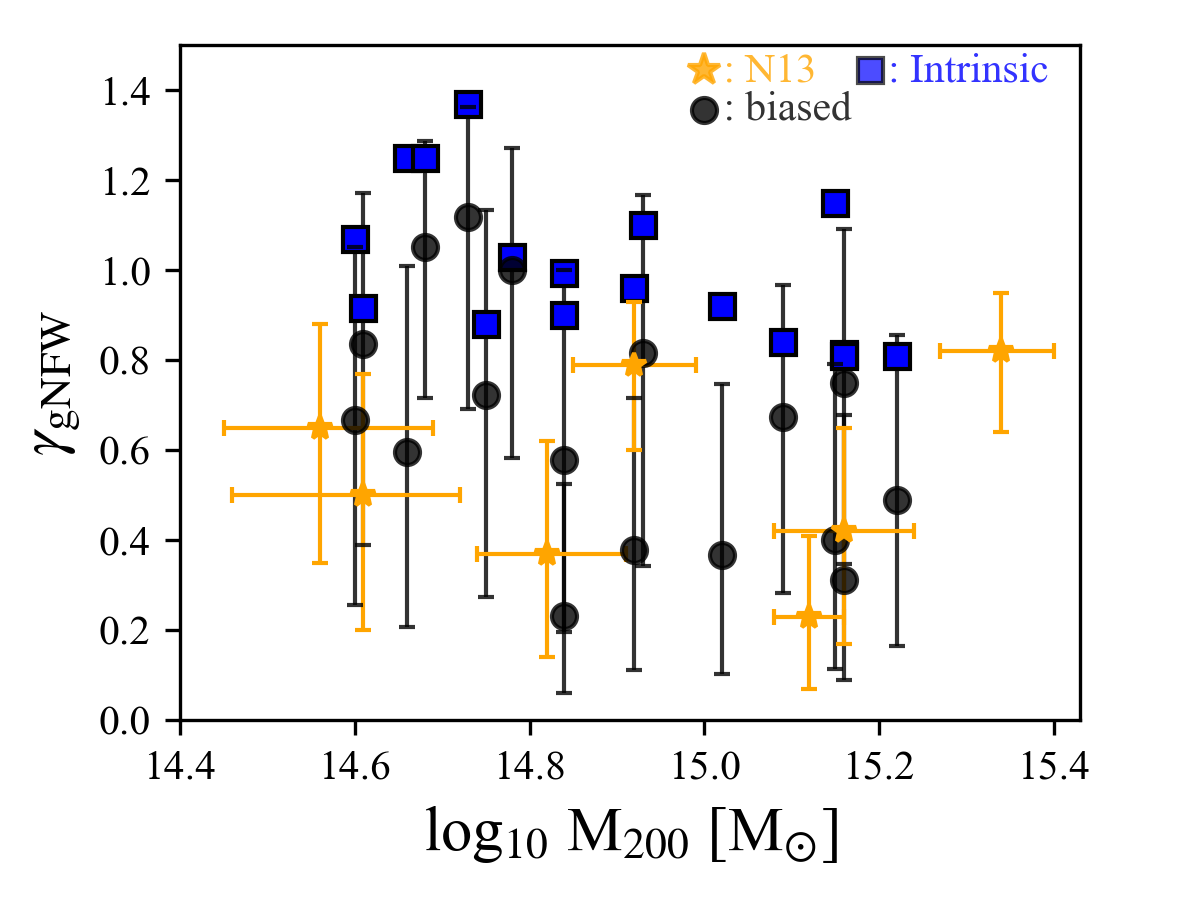}
	\caption{Comparison of the values of $\gamma_{\rm gNFW}$
          inferred from biased weak lensing inputs (black points),
          derived by  \citet{Newman2013b} (orange stars) and the
          actual values for C-EAGLE clusters (blue squares).}
\label{fig:bias_gamma}
\end{figure}

  The best-fit values of $\gamma_{\rm gNFW}$ are shown as black points
  in Fig.~\ref{fig:bias_gamma}. As may be seen, these slopes,  
  derived assuming artificially biased weak lensing inputs, are much
  smaller than the true values shown in blue. They are, in fact, quite
  comparable to the results of  \citet{Newman2013b}. Of
  course, we do not claim that the latter are biased but our
  conclusions point to one possible way in which the discrepancy
  between the results of \citet{Newman2013b} and our simulations might be
  resolved. 

\begin{table}
	\centering
	\caption{Comparison amongst the different lensing measurements of 
          the NFW scale radius (in kiloparsecs) for three clusters, MS2137,
          A383 and A611, obtained  by \citet{Newman2013a},
          \citet{Merten2015} and \citet{Umetsu2016} (denoted as N13,
          M15 and U16, respectively). For
          convenience, we adopt $h=0.7$} 
         \label{tab:NFW_rs}
    \renewcommand{\arraystretch}{1.5}
	\begin{tabular}{cccc}
		\hline
		\hline
		&  MS2137 & A383 & A611\\
		\hline
		N13 & 119$^{+49}_{-32}$ & 260$^{+59}_{-45}$ & 317$^{+57}_{-47}$ \\
		M15 & 686$^{+71}_{-71}$ & 471$^{+57}_{-57}$ & 586$^{+86}_{-86}$ \\
		U16 & 800$^{+450}_{-450}$ & 310$^{+130}_{-130}$ & 570$^{+210}_{-210}$ \\
		\hline
		\hline
	\end{tabular}
	
\end{table}
%


%

\subsection{Robustness to model assumptions}

In this section we consider the effect of various model assumptions on
the estimates of the inner dark matter slope.

\subsubsection{Mass-to-light ratio}\label{sec:masslight}

In the preceding analysis we made use of our fiducial mock data in
which a constant mass-to-light ratio was assumed when generating the
surface brightness map of the central galaxy. However, this may not
apply in the real Universe. To explore the sensitivity of our results
to this assumption, we built another set of mocks, this time using the
$r$-band luminosity calculated with the photometric method of
\citet{Trayford2015}. We performed the same analysis on this new set
of mocks, still assuming a constant $M^*/L$. The difference between
these results and those from our fiducial model reflects the
uncertainties introduced by the simple assumption of constant $M^*/L$.

\begin{figure}
 	\includegraphics[width=0.5\textwidth]{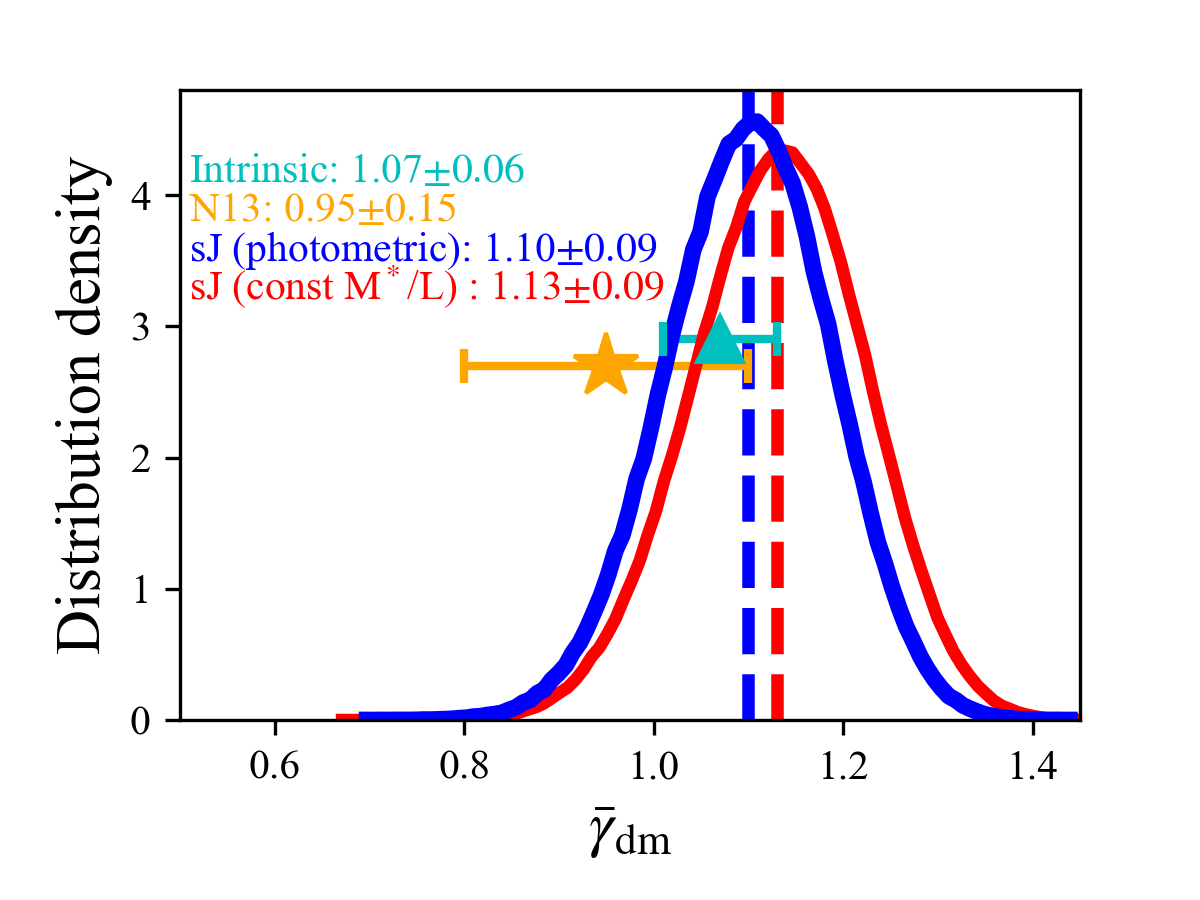}
	\caption{Comparison of the joint constraints on the mean
          $\bar{\gamma}_{\rm dm}$ (at 44 kpc) from mock data
          constructed either assuming a constant ${M^*/L}$ (red) or the
          photometric model of \citet{Trayford2015} (blue).  Results
          are shown for sJ + lensing modelling, assuming in both cases
          that ${M^*/L}$ is constant. The yellow symbol and error bar
          show the observational result of \citet{Newman2013b}, while
          the cyan symbol and error bar correspond to the true C-EAGLE
          result. The values quoted in the legend are the most
          probable values of the mean $\bar{\gamma}_{\rm dm}$ derived
          from the corresponding test.}
\label{fig:mlgamma}
\end{figure}

In Fig.~\ref{fig:mlgamma} we compare the joint constraints on the mean
$\bar{\gamma}_{\rm dm}$ (at 44 kpc) from our mock data using the sJ + lensing
model, assuming a constant $M^*/L$, for both, mocks constructed making
this same assumption and mocks constructed using the photometric model
of \citet{Trayford2015}.  The inferred mean $\bar{\gamma}_{\rm dm}$ in
the latter case is only about 3\% smaller than in the standard
case. These results indicate that the assumption of a constant $M^*/L$
is reasonable for the analysis of the inner dark matter density
profiles in these massive clusters.

\subsubsection{Shape of the central galaxy}

\begin{figure}
	\includegraphics[width=0.52\textwidth]{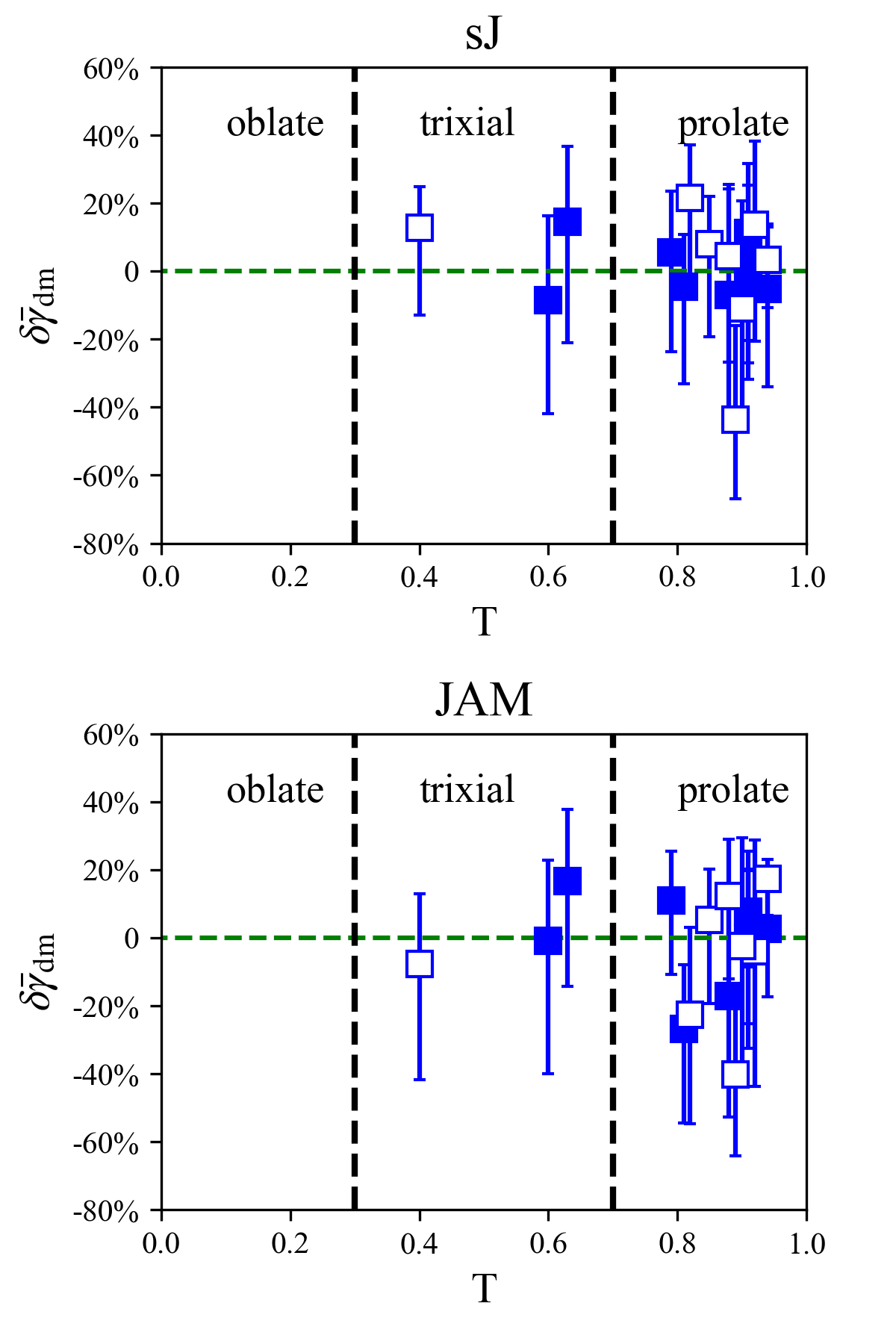}
	\caption{$\delta\bar{\gamma}_{\rm dm}$ as a function of the
          triaxiality parameter. The solid squares show relaxed
          clusters and the empty squares unrelaxed
          clusters.} \label{fig:shape}
\end{figure}

When modeling the central stellar dynamics by solving the Jeans
equations, we assumed the galaxy to have either a spherical or an
oblate shape. A spherical shape for the central galaxy is assumed in
the sJ model, while an oblate shape is assumed in the fiducial JAM
model. Although this oblateness assumption is valid for most early
type galaxies, it does not apply to the most massive ones
\citep[e.g.][]{Hongyu2016,Hongyu2017}. To be consistent with previous
analyses, we assumed oblateness in our application of the JAM. In the
upper and lower panel of Fig.~\ref{fig:shape} we show the error in the
inferred mass-weighted slope of the dark matter density profile,
$\delta\bar{\gamma}_{\rm dm}=(\bar{\gamma}_{\rm dm}^{'} -
\bar{\gamma}_{\rm dm})/\bar{\gamma}_{\rm dm}$ (where, as
before, $\bar{\gamma}^{'}_{\rm dm}$ denotes the best-fit value and
$\bar{\gamma}_{\rm dm}$ the true value), as a function of the
triaxiality parameter, $T\equiv\frac{a^2-b^2}{a^2-c^2}$
\citep{Binney2008} for both the sJ and JAM models.

We compute the triaxiality parameter of the galaxy using the
reduced inertia tensor defined as:
\begin{equation}
I_{ij, k+1} = \frac{\sum\limits_{n}M_{n}x_{i, n}x_{j, n}/r_{n, k}^{2}}{\sum\limits_{n}M_n}\:,
\end{equation}
where $i,j \in \{x, y, z\}$ and the summation is over the stars within
25 kpc, (which is slightly larger than the region with 
  kinematical data and around 2 R$_e$ for our sample. Here, $r_{n, k}$ is
defined as the $k$-th iteration value of the radius,
\begin{equation}\label{equ:shape}
r_{n,k} = \sqrt{x_n^2+y_n^2/q^2+z_n^2/s^2}\:,
\end{equation}
where $q=b/a$ and $s=c/a$ (assuming the lengths of the three major
axes are $a$, $b$, $c$ and $a\geq b\geq c$) are the square root of the
ratios of the reduced inertia tensor eigenvalues. We iteratively
calculate the reduced inertia tensor and the values of $q$ and $s$,
deriving the triaxiality parameter from the stable $q$ and $s$ values.

If $s\ge0.9$, then the galaxy is close to spherical and, if $s\le0.9$,
we can classify the shape into three categories: oblate for $T\le0.3$,
prolate for $T\ge0.7$ and triaxial in between. All of our clusters have $s\le0.9$. From
Fig.~\ref{fig:shape}, we find that most of the cluster central
galaxies have a prolate shape. Interestingly, although the shapes are not consistent with the assumption of the JAM or the sJ model, we
do not find a correlation between the accuracy of the estimate of
$\bar{\gamma}_{\rm dm}$ and the triaxiality parameter.

\begin{figure}
	\includegraphics[width=0.5\textwidth]{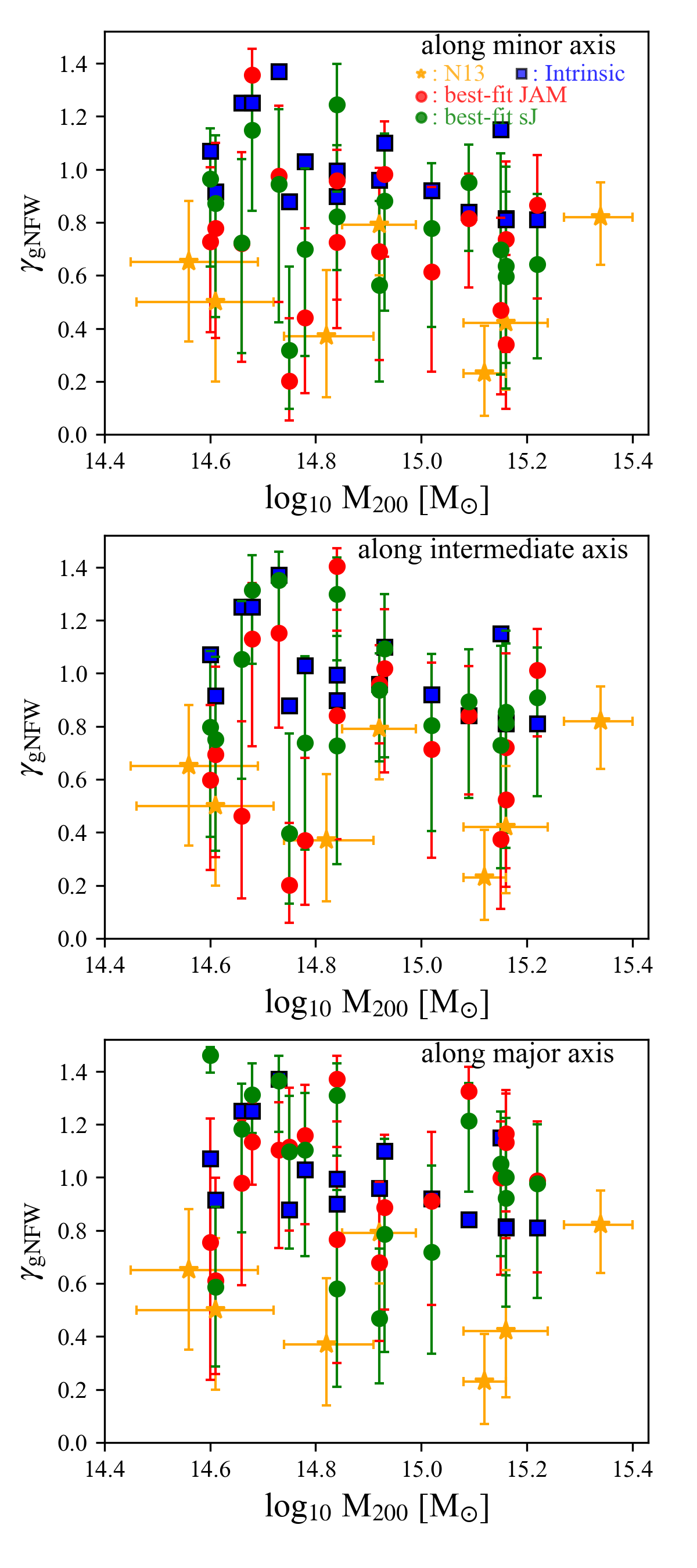}
	\caption{Values of $\gamma_{\rm gNFW}$ for C-EAGLE clusters
          viewed from different directions. The upper, middle and
          lower panels show results when viewing the central galaxies
          along the minor, intermediate and major axis,
          respectively. Blue squares are the true C-EAGLE values and
          the yellow stars the measured values of
          \citet{Newman2013b}. Red and green circles are results from
          the JAM and sJ model
          respectively.} \label{fig:rotation_slope}
\end{figure}

To explore further the model dependence on the galaxy shape, we rotate
all of our galaxies in three different directions, so that the
line-of-sight direction is aligned with the major, intermediate and
minor axes respectively, and repeat the kinematics + lensing
analysis. We show the best-fit $\gamma_{\rm gNFW}$ in different
directions as a function of M$_{200}$ in
Fig.~\ref{fig:rotation_slope}. We see that for both models looking
along intermediate and minor axes gives similar $\gamma_{\rm gNFW}$
distributions, while looking along the longest axis gives larger
best-fit values of $\gamma_{\rm gNFW}$ than for the two other
directions. The probability of drawing 7 asymptotic slope values from
their posterior distributions with mean value lying within the
1$\sigma$ range of the observational result ($0.50\pm0.13$) are
17.3\%, 5.7\% and 1.5\% when viewing along minor, intermediate and
major axes, respectively.

\subsubsection{Velocity anisotropy}

We also test the dependence of the model on the velocity
anisotropy. \citet{Schaller2015b} suggested that the discrepancy
between the dark matter density profile slopes in the observed
clusters and in the EAGLE simulations might be due to incorrect
assumptions for the velocity anisotropy parameters. In the sJ
modeling, the velocity anisotropy is assumed to be zero, while in JAM
it is assumed to be constant in the $z$ cylindrical coordinate.

\begin{figure}
	\includegraphics[width=0.52\textwidth]{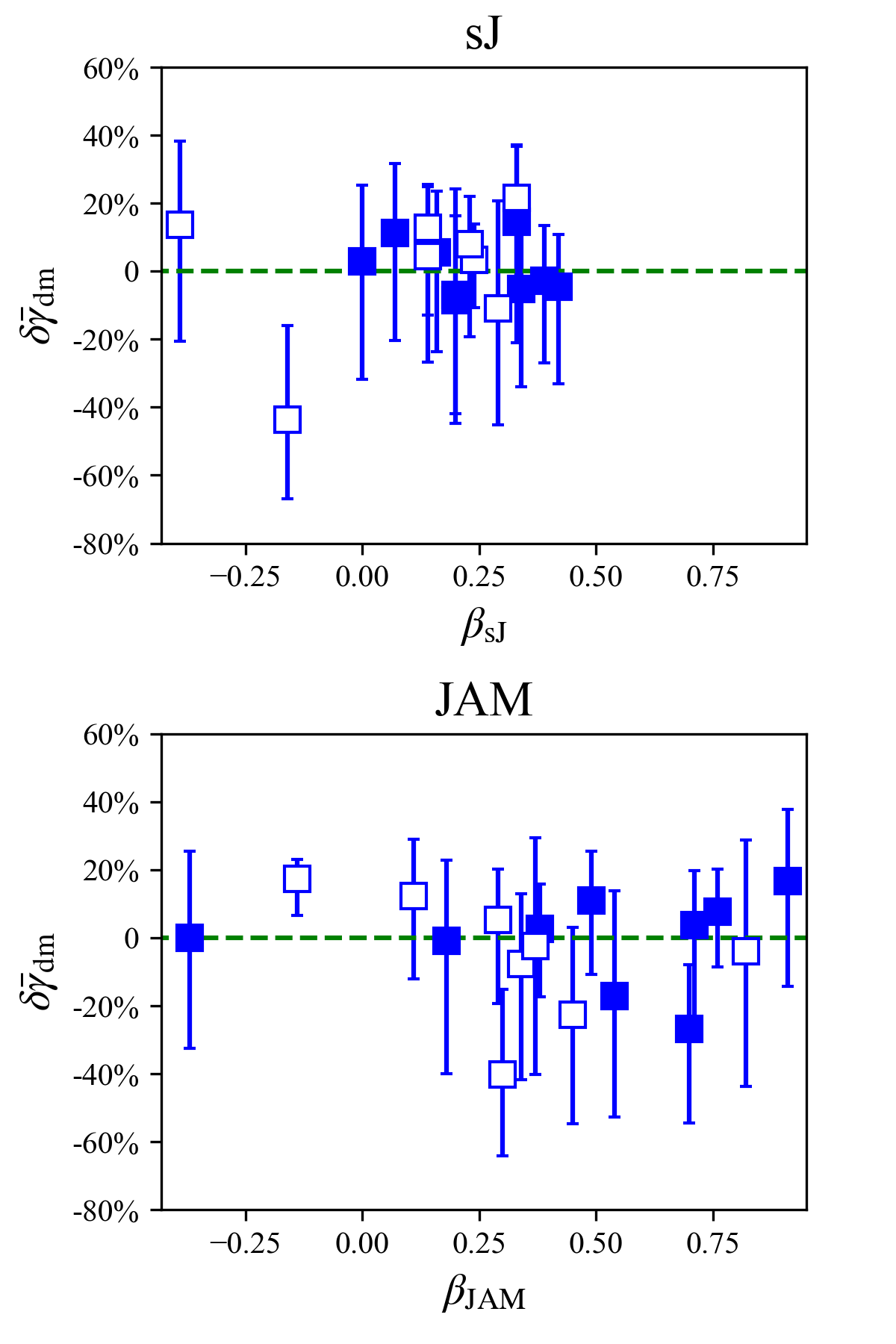}
        \caption{$\delta\bar{\gamma}_{\rm dm}$ as a function of the
          anisotropy parameter, $\beta$. The solid squares show
          relaxed clusters and the empty squares unrelaxed
          clusters.} \label{fig:betaz}
\end{figure}

In Fig.~\ref{fig:betaz}  we plot the error in the estimates of
$\bar{\gamma}_{\rm dm}$ as a function of the anisotropy parameter,
$\beta$, of the C-EAGLE clusters for both the sJ and JAM 
cases. The anisotropy in cylindrical coordinates, $\beta_{\rm JAM}$
($\beta_z$), is computed as
\begin{equation}
\beta_{\rm JAM} = 1 - \frac{\overline{v_z^2}}{\overline{v_R^2}}\:,
\end{equation}
where the $z$-axis is aligned with the minor axis of the galaxy. The
anisotropy parameter for the spherical Jeans model is computed as:
\begin{equation}
\beta_{\rm sJ} = 1 - \frac{\overline{v_\theta^2}}{\overline{v_r^2}}\:,
\end{equation}
where $\overline{v_\theta^2}=\overline{v_\phi^2}$. There is 
significant galaxy to galaxy scatter in the anisotropy parameter value
but we do not find a significant trend of 
$\delta\bar{\gamma}_{\rm dm}$ with $\beta$.

\section{Summary and Discussion}

We have investigated the accuracy of techniques for inferring the
inner density profiles of massive galaxy clusters from a combination
of stellar kinematics and gravitational lensing data. We constructed
mock datasets from 17 clusters in the C-EAGLE hydrodynamical
simulations \citep{Barnes2017,Bahe2017}, whose masses are comparable
to those of the seven clusters studied by \citet{Newman2013a} (with a
mean M$_{\rm 200}\sim 1\times10^{15}$ $\rm{M}_{\odot}$). We performed
a stellar dynamical and lensing analysis on the mock datasets. For the
former we used two different methods: the spherical Jeans model, which
was the method used by \citet{Newman2013a,Newman2013b}, and the Jeans
anistropic model. Our findings can be summarized as follows:

\begin{itemize}

\item The values of the inner asymptotic slope of a ``generalized''
  NFW density profile, $\gamma_{\rm gNFW}$, estimated using the
  kinematics + lensing analysis on the mock data agree reasonably
  well with the input values indicating that, in principle, the method
  is accurate and unbiased.

\item The dark matter asymptotic density slopes, $\gamma_{\rm gNFW}$,
  of massive C-EAGLE clusters are steeper than those inferred by
  \citet{Newman2013a,Newman2013b} for the observed clusters.  The 
  C-EAGLE clusters have $\gamma_{\rm gNFW} \sim 1$, whereas
  \citet{Newman2013b} find $\gamma_{\rm gNFW} \sim 0.5$ for their
  clusters, as shown in Fig.~\ref{fig:gamma_comparison}. 


\item The inner density profile can also be characterized by the mean
  mass-weighted dark matter density slope, $\bar{\gamma}_{\rm dm}$,
  averaged within the effective optical radius of the central
  galaxy. To compare our results with observations, we
    derive this average slope within 44 kpc, which is approximately
    the effective radius of the clusters of
    \citet{Newman2013a}. Taking errors into account, the average
    slopes from C-EAGLE clusters agree with the observed ones (see
    Fig.~\ref{fig:ngMgamma}).

\item The different conclusions reached when using the
    two different measures of inner dark matter density profile slope
    can be traced back to different values of the characteristic halo
    radius, $r_s$, in the C-EAGLE and observed clusters. The values of
    $r_s$ inferred for the \citet{Newman2013a} sample are
    significantly smaller than the values for the clusters in the
    simulations (see Fig.~\ref{fig:rs_plot}). The smaller the $r_s$,
    the faster the dark matter density slope varies within the
    effective radius and thus the larger the difference between the
    asymptotic and mass-weighted values.

    {\item We find that there is a strong degeneracy
      between the asymptotic gNFW slope, $\gamma_{\rm gNFW}$, and the
      scale radius, $r_s$ (or, equivalently, the scale density,
      $\rho_{\rm s}$; see Figs.~\ref{fig:mcmc_sJ}
      and~\ref{fig:halo13}). As a result, the gravitational lensing
      data which directly probe the cluster mass distribution at
      large distances also play a role in constraining the inner
      profiles. To assess the importance of lensing data, we repeated
      our analysis of the C-EAGLE clusters in two ways. Firstly, we
      ignored lensing and used only stellar kinematical data. We found
      that, in this case, the dark matter density slopes are
      significantly underestimated (see
      Fig.~\ref{fig:gamma_M200_nl}). This is probably because, as
      shown in Fig.~\ref{fig:rms_plot}, not including constraints from
      lensing loses the anchor point in the outer regions of the
      cluster and a nearly constant density dark matter core is then
      preferred to account for the steeply raising observed stellar
      velocity dispersion profile (which otherwise the stellar
      dynamical models considered here would have difficulty
      matching).

      Secondly, we kept the stellar kinematical and strong lensing
      mock data unchanged, but artificially biased the weak lensing
      mock data to correspond to a profile with a 50\% smaller value
      of the NFW $r_s$. We found that, in this case, the best-fit
      $\gamma_{\rm gNFW}$ values are significantly underestimated and
      are, in fact, quite comparable with the values estimated by
      \citet{Newman2013a,Newman2013b}. We noted that for three
      clusters, the NFW scale radii measured by \citet{Newman2013a}
      are much smaller than the more recent measurements carried out
      by \citet{Merten2015} and \citet{Umetsu2016}. Based on these
      tests, we suggest that stellar kinematical data combined with
      lensing measurements from the more recent observations would
      alleviate the discrepancy between the observed dark matter
      density slopes and the theoretical predictions. We also note,
      however, that the haloes of three of the observed clusters have
      scale radii similar to those in the simulations (lower panel of
      Fig.~\ref{fig:rs_plot}), so biased lensing results may not be
      the whole story behind the tension.}
  
\item We also applied our sJ + lensing and JAM + lensing analyses to
  clusters viewed from their minor, intermediate and major axes. We
  found that the best-fit $\gamma_{\rm gNFW}$ tends to be larger than
  the true value when the cluster is viewed from the direction of the
  major axis. If the observed sample were biased in this way, the
  discrepancy with the C-EAGLE clusters would be even larger. 

\item We tested the robustness of the method to the assumptions of a
  constant stellar mass-to-light ratio and an isotropic velocity
  anisotropy and found the method to be fairly insensitive to these
  assumptions.

  In summary, while according to one measurement (the
    mean inner slope of the dark matter density profile) the
    observational data agree with the simulations, according to
    another (the asymptotic slope) they do not. These two measures
    differ in the way they weight different regions of the mass
    distribution in the cluster. The asymptotic slopes are
    extrapolations that rely on the innermost data points whereas the
    mean slopes may be more robust. The inferred asymptotic slopes are
    degenerate with the scale radius (or the scale density) of the
    halo and thus they can be strongly affected by the lensing data; a
    poor or biased measurement of the lensing constraints can lead to
    significantly smaller asymptotic slopes. Thus, although some
    tension between the simulations and the data remains, this does
    not necessarily imply a fatal inconsistency between the two.

\end{itemize}

\label{sec:sum}
\section*{Acknowledgements}
We thank Mathilde Jauzac and Andrew Robertson for helpful
  suggestions. We are also very grateful to an anonymous referee whose perceptive comments
led to significant improvements to our paper. We acknowledge National Natural Science Foundation of
China (Nos. 11773032), and National Key Program for Science and
Technology Research and Development of China (2017YFB0203300). RL is
supported by NAOC Nebula Talents Program.

This work was supported by the CSF's European Research Council (ERC)
Advanced Investigator grant DMIDAS (GA 786910) and the Consolidated
Grant for Astronomy at Durham (ST/L00075X/1). It used the DiRAC Data
Centric system at Durham University, operated by the Institute for
Computational Cosmology on behalf of the STFC DiRAC HPC Facility
(\url{www.dirac.ac.uk}). This equipment was funded by BIS National
E-infrastructure capital grant ST/K00042X/1, STFC capital grants
ST/H008519/1 and ST/K00087X/1, STFC DiRAC Operations grant
ST/K003267/1 and Durham University. DiRAC is part of the National
E-Infrastructure. MS is supported by VENI grant 639.041.749. YMB acknowledges funding from the EU Horizon 2020
research and innovation programme under Marie Sk{\l}odowska-Curie
grant agreement 747645 (ClusterGal) and the Netherlands Organisation
for Scientific Research (NWO) through VENI grant 016.183.011. The
C-EAGLE simulations were in part performed on the German federal
maximum performance computer~``HazelHen'' at the maximum performance
computing centre Stuttgart (HLRS), under project GCS-HYDA / ID 44067
financed through the large-scale project~``Hydrangea'' of the Gauss
Center for Supercomputing. Further simulations were performed at the
Max Planck Computing and Data Facility in Garching,
Germany. CDV acknowledges financial support from the Spanish Ministry of Economy and Competitiveness (MINECO) through grants AYA2014-58308 and RYC-2015-1807.

\section*{Data availability}
The data underlying this article will be shared on reasonable request to the corresponding author.

\bibliography{reference}

\appendix
\section{Tables for key recovered parameters}
We list best-fit and true values for six key parameters in Table~\ref{tab:results} and Table~\ref{tab:results_sJ} for sJ and JAM + lensing analysis, respectively.

\begin{table*}
	\centering
	\caption{Comparison between true and best-fit parameters
          obtained from sJ plus lensing analysis. The best-fit and
          true values are denoted with superscript "R" and "T"
          respectively. $\bar{\gamma}_{\rm tot}$ and $\bar{\gamma}_{\rm dm}$
          are the mass-weighted slope of total density profiles and dark matter
          density profiles and $\gamma_{\rm gNFW}$ is the asymptotic
          dark matter density slope. M$^*$ and M$_{\rm tot}$ are the
          stellar and total mass enclosed within the stellar effective radius,
          R$_{\rm e}$. The unit of mass is M$_\odot$. f$_{\rm dm}$ is the
          dark matter mass fraction witnin R$_{\rm e}$. Errors are calculated
          as 84 and 16 percentiles.}\label{tab:results}
    \renewcommand{\arraystretch}{1.5}
	\begin{tabular}{|c|cccccccccccc|}
	
		\hline
		\hline
		& $\bar{\gamma}^{\rm T}_{\rm tot}$ & $\bar{\gamma}^{\rm R}_{\rm tot}$ & $\bar{\gamma}^{\rm T}_{\rm dm}$ & $\bar{\gamma}^{\rm R}_{\rm dm}$ & $\gamma^{\rm T}_{\rm gNFW}$ & $\gamma^{\rm R}_{\rm gNFW}$ & log$_{10}$ M$^{*\rm{T}}$ & log$_{10}$ M$^{*\rm{R}}$ & log$_{10}$ M$^{\rm T}_{\rm tot}$ & log$_{10}$ M$^{\rm R}_{\rm tot}$ & f$^{\rm T}_{\rm dm}$ & f$^{\rm R}_{\rm dm}$ \\
		\hline
		CE-12 & 1.79 & 1.63$^{+0.05}_{-0.05}$ & 1.15 & 1.09$^{+0.21}_{-0.33}$ & 1.07 & 1.03$^{+0.24}_{-0.42}$ & 11.62 & 11.76$^{+0.09}_{-0.13}$ & 11.91 & 12.07$^{+0.03}_{-0.04}$ & 0.49 & 0.50$^{+0.13}_{-0.13}$ \\
		CE-13 & 1.69 & 1.65$^{+0.06}_{-0.06}$ & 0.98 & 1.09$^{+0.20}_{-0.31}$ & 0.92 & 1.01$^{+0.24}_{-0.41}$ & 11.63 & 11.69$^{+0.10}_{-0.13}$ & 11.91 & 12.01$^{+0.02}_{-0.03}$ & 0.48 & 0.52$^{+0.13}_{-0.14}$ \\
		CE-14 & 1.80 & 1.76$^{+0.06}_{-0.05}$ & 0.97 & 1.18$^{+0.15}_{-0.27}$ & 1.25 & 1.15$^{+0.16}_{-0.32}$ & 11.58 & 11.59$^{+0.08}_{-0.10}$ & 11.81 & 11.83$^{+0.02}_{-0.03}$ & 0.41 & 0.42$^{+0.12}_{-0.12}$ \\
		CE-15 & 2.05 & 1.86$^{+0.05}_{-0.05}$ & 1.17 & 1.32$^{+0.14}_{-0.30}$ & 1.37 & 1.28$^{+0.16}_{-0.38}$ & 11.56 & 11.57$^{+0.09}_{-0.09}$ & 11.73 & 11.79$^{+0.03}_{-0.03}$ & 0.32 & 0.41$^{+0.10}_{-0.13}$ \\
		CE-16 & 1.76 & 1.78$^{+0.09}_{-0.08}$ & 1.07 & 1.10$^{+0.24}_{-0.37}$ & 0.88 & 1.04$^{+0.26}_{-0.45}$ & 11.69 & 11.99$^{+0.11}_{-0.15}$ & 11.94 & 12.21$^{+0.03}_{-0.04}$ & 0.44 & 0.40$^{+0.17}_{-0.15}$ \\
		CE-17 & 1.75 & 1.63$^{+0.07}_{-0.07}$ & 1.22 & 1.26$^{+0.13}_{-0.17}$ & 1.25 & 1.07$^{+0.23}_{-0.27}$ & 11.63 & 11.72$^{+0.11}_{-0.12}$ & 11.95 & 12.16$^{+0.02}_{-0.02}$ & 0.53 & 0.64$^{+0.09}_{-0.10}$ \\
		CE-18 & 1.65 & 1.58$^{+0.06}_{-0.06}$ & 1.05 & 0.96$^{+0.26}_{-0.35}$ & 0.90 & 0.85$^{+0.32}_{-0.49}$ & 11.73 & 11.83$^{+0.09}_{-0.14}$ & 12.02 & 12.11$^{+0.02}_{-0.02}$ & 0.47 & 0.48$^{+0.15}_{-0.13}$ \\
		CE-19 & 1.95 & 1.54$^{+0.08}_{-0.07}$ & 1.12 & 0.63$^{+0.31}_{-0.26}$ & 0.99 & 0.50$^{+0.38}_{-0.35}$ & 11.65 & 11.71$^{+0.06}_{-0.08}$ & 11.92 & 11.94$^{+0.04}_{-0.04}$ & 0.47 & 0.41$^{+0.13}_{-0.10}$ \\
		CE-20 & 1.90 & 1.71$^{+0.05}_{-0.05}$ & 1.14 & 1.23$^{+0.16}_{-0.31}$ & 1.03 & 1.18$^{+0.19}_{-0.43}$ & 11.72 & 11.88$^{+0.10}_{-0.14}$ & 12.01 & 12.22$^{+0.02}_{-0.02}$ & 0.48 & 0.54$^{+0.12}_{-0.13}$ \\
		CE-21 & 1.88 & 1.79$^{+0.06}_{-0.05}$ & 1.10 & 1.16$^{+0.20}_{-0.32}$ & 1.10 & 1.10$^{+0.23}_{-0.43}$ & 11.86 & 11.97$^{+0.08}_{-0.11}$ & 12.08 & 12.21$^{+0.02}_{-0.02}$ & 0.40 & 0.42$^{+0.13}_{-0.12}$ \\
		CE-22 & 1.56 & 1.55$^{+0.06}_{-0.06}$ & 0.87 & 0.81$^{+0.27}_{-0.33}$ & 0.92 & 0.74$^{+0.30}_{-0.44}$ & 11.81 & 11.87$^{+0.07}_{-0.10}$ & 12.10 & 12.11$^{+0.03}_{-0.03}$ & 0.48 & 0.41$^{+0.14}_{-0.11}$ \\
		CE-23 & 1.47 & 1.48$^{+0.07}_{-0.06}$ & 0.73 & 0.83$^{+0.18}_{-0.25}$ & 0.96 & 0.77$^{+0.20}_{-0.29}$ & 11.62 & 11.73$^{+0.07}_{-0.10}$ & 11.93 & 12.01$^{+0.04}_{-0.04}$ & 0.51 & 0.47$^{+0.13}_{-0.11}$ \\
		CE-24 & 1.63 & 1.60$^{+0.06}_{-0.06}$ & 1.04 & 1.01$^{+0.17}_{-0.25}$ & 0.84 & 0.97$^{+0.18}_{-0.31}$ & 11.76 & 11.90$^{+0.08}_{-0.12}$ & 12.08 & 12.17$^{+0.02}_{-0.03}$ & 0.52 & 0.46$^{+0.13}_{-0.12}$ \\
		CE-25 & 1.74 & 1.64$^{+0.05}_{-0.05}$ & 1.12 & 1.07$^{+0.17}_{-0.32}$ & 1.15 & 1.01$^{+0.20}_{-0.44}$ & 11.85 & 11.89$^{+0.08}_{-0.11}$ & 12.11 & 12.14$^{+0.02}_{-0.02}$ & 0.45 & 0.44$^{+0.12}_{-0.12}$ \\
		CE-26 & 1.63 & 1.51$^{+0.07}_{-0.07}$ & 0.82 & 0.73$^{+0.26}_{-0.28}$ & 0.82 & 0.62$^{+0.33}_{-0.38}$ & 11.90 & 11.89$^{+0.07}_{-0.10}$ & 12.18 & 12.15$^{+0.03}_{-0.03}$ & 0.47 & 0.44$^{+0.12}_{-0.10}$ \\
		CE-27 & 1.40 & 1.45$^{+0.06}_{-0.07}$ & 0.90 & 1.03$^{+0.20}_{-0.32}$ & 0.81 & 0.96$^{+0.24}_{-0.42}$ & 11.56 & 11.81$^{+0.15}_{-0.23}$ & 12.01 & 12.20$^{+0.02}_{-0.03}$ & 0.64 & 0.59$^{+0.17}_{-0.17}$ \\
		CE-28 & 1.53 & 1.41$^{+0.06}_{-0.07}$ & 0.90 & 0.94$^{+0.19}_{-0.28}$ & 0.81 & 0.85$^{+0.23}_{-0.38}$ & 11.81 & 11.96$^{+0.09}_{-0.11}$ & 12.16 & 12.36$^{+0.03}_{-0.03}$ & 0.55 & 0.60$^{+0.10}_{-0.11}$ \\
		\hline
		\hline
	\end{tabular}
\end{table*}

\begin{table*}
	\centering
	\caption{Comparison between true and best-fit parameters
          obtained from JAM plus lensing
          analysis. Notations are the same as Table
          \ref{tab:results}.}\label{tab:results_sJ}
    \renewcommand{\arraystretch}{1.5}
	\begin{tabular}{|c|cccccccccccc|}
		\hline
		\hline
		
		& $\bar{\gamma}^{\rm T}_{\rm tot}$ & $\bar{\gamma}^{\rm R}_{\rm tot}$ & $\bar{\gamma}^{\rm T}_{\rm dm}$ & $\bar{\gamma}^{\rm R}_{\rm dm}$ & $\gamma^{\rm T}_{\rm gNFW}$ & $\gamma^{\rm R}_{\rm gNFW}$ & log$_{10}$ M$^{*\rm{T}}$ & log$_{10}$ M$^{*\rm{R}}$ & log$_{10}$ M$^{\rm T}_{\rm tot}$ & log$_{10}$ M$^{\rm R}_{\rm tot}$ & f$^{\rm T}_{\rm dm}$ & f$^{\rm R}_{\rm dm}$ \\
		\hline
		
		CE-12 & 1.79 & 1.70$^{+0.06}_{-0.06}$ & 1.15 & 1.18$^{+0.15}_{-0.23}$ & 1.07 & 1.13$^{+0.17}_{-0.30}$ & 11.62 & 11.65$^{+0.08}_{-0.10}$ & 11.91 & 12.01$^{+0.03}_{-0.04}$ & 0.49 & 0.56$^{+0.11}_{-0.11}$ \\
		CE-13 & 1.69 & 1.74$^{+0.08}_{-0.07}$ & 0.98 & 0.98$^{+0.25}_{-0.32}$ & 0.92 & 0.90$^{+0.31}_{-0.42}$ & 11.63 & 11.68$^{+0.07}_{-0.10}$ & 11.91 & 11.94$^{+0.04}_{-0.04}$ & 0.48 & 0.45$^{+0.14}_{-0.12}$ \\
		CE-14 & 1.80 & 1.98$^{+0.08}_{-0.08}$ & 0.97 & 0.75$^{+0.25}_{-0.31}$ & 1.25 & 0.69$^{+0.31}_{-0.40}$ & 11.58 & 11.61$^{+0.04}_{-0.05}$ & 11.81 & 11.73$^{+0.03}_{-0.03}$ & 0.41 & 0.23$^{+0.09}_{-0.07}$ \\
		CE-15 & 2.05 & 2.02$^{+0.09}_{-0.08}$ & 1.17 & 1.08$^{+0.24}_{-0.40}$ & 1.37 & 1.03$^{+0.26}_{-0.46}$ & 11.56 & 11.54$^{+0.07}_{-0.08}$ & 11.73 & 11.69$^{+0.03}_{-0.04}$ & 0.32 & 0.28$^{+0.13}_{-0.11}$ \\
		CE-16 & 1.76 & 1.76$^{+0.14}_{-0.10}$ & 1.07 & 1.11$^{+0.17}_{-0.31}$ & 0.88 & 1.07$^{+0.20}_{-0.36}$ & 11.69 & 11.77$^{+0.14}_{-0.15}$ & 11.94 & 12.07$^{+0.04}_{-0.04}$ & 0.44 & 0.50$^{+0.14}_{-0.18}$ \\
		CE-17 & 1.75 & 1.71$^{+0.05}_{-0.05}$ & 1.22 & 1.43$^{+0.07}_{-0.13}$ & 1.25 & 1.39$^{+0.08}_{-0.16}$ & 11.63 & 11.62$^{+0.11}_{-0.12}$ & 11.95 & 12.09$^{+0.02}_{-0.02}$ & 0.53 & 0.67$^{+0.08}_{-0.10}$ \\
		CE-18 & 1.65 & 1.76$^{+0.10}_{-0.09}$ & 1.05 & 1.04$^{+0.25}_{-0.41}$ & 0.90 & 0.98$^{+0.29}_{-0.53}$ & 11.73 & 11.84$^{+0.09}_{-0.16}$ & 12.02 & 12.10$^{+0.03}_{-0.03}$ & 0.47 & 0.44$^{+0.18}_{-0.16}$ \\
		CE-19 & 1.95 & 1.69$^{+0.10}_{-0.09}$ & 1.12 & 0.67$^{+0.28}_{-0.27}$ & 0.99 & 0.55$^{+0.34}_{-0.34}$ & 11.65 & 11.67$^{+0.06}_{-0.07}$ & 11.92 & 11.91$^{+0.05}_{-0.05}$ & 0.47 & 0.42$^{+0.12}_{-0.10}$ \\
		CE-20 & 1.90 & 1.79$^{+0.08}_{-0.07}$ & 1.14 & 1.20$^{+0.17}_{-0.28}$ & 1.03 & 1.16$^{+0.18}_{-0.34}$ & 11.72 & 11.84$^{+0.08}_{-0.11}$ & 12.01 & 12.15$^{+0.04}_{-0.04}$ & 0.48 & 0.52$^{+0.13}_{-0.13}$ \\
		CE-21 & 1.88 & 1.84$^{+0.07}_{-0.06}$ & 1.10 & 1.22$^{+0.16}_{-0.24}$ & 1.10 & 1.18$^{+0.18}_{-0.30}$ & 11.86 & 11.88$^{+0.07}_{-0.09}$ & 12.08 & 12.15$^{+0.04}_{-0.04}$ & 0.40 & 0.46$^{+0.12}_{-0.11}$ \\
		CE-22 & 1.56 & 1.77$^{+0.08}_{-0.08}$ & 0.87 & 0.72$^{+0.27}_{-0.31}$ & 0.92 & 0.63$^{+0.31}_{-0.40}$ & 11.81 & 11.96$^{+0.04}_{-0.06}$ & 12.10 & 12.12$^{+0.03}_{-0.03}$ & 0.48 & 0.29$^{+0.12}_{-0.09}$ \\
		CE-23 & 1.47 & 1.66$^{+0.09}_{-0.09}$ & 0.73 & 0.70$^{+0.24}_{-0.29}$ & 0.96 & 0.64$^{+0.26}_{-0.36}$ & 11.62 & 11.79$^{+0.05}_{-0.06}$ & 11.93 & 11.99$^{+0.05}_{-0.05}$ & 0.51 & 0.35$^{+0.13}_{-0.10}$ \\
		CE-24 & 1.63 & 1.68$^{+0.07}_{-0.06}$ & 1.04 & 1.12$^{+0.13}_{-0.17}$ & 0.84 & 1.09$^{+0.14}_{-0.20}$ & 11.76 & 11.79$^{+0.07}_{-0.10}$ & 12.08 & 12.13$^{+0.03}_{-0.03}$ & 0.52 & 0.54$^{+0.11}_{-0.11}$ \\
		CE-25 & 1.74 & 1.89$^{+0.08}_{-0.08}$ & 1.12 & 0.82$^{+0.21}_{-0.31}$ & 1.15 & 0.76$^{+0.24}_{-0.40}$ & 11.85 & 11.96$^{+0.05}_{-0.06}$ & 12.11 & 12.09$^{+0.02}_{-0.03}$ & 0.45 & 0.26$^{+0.09}_{-0.08}$ \\
		CE-26 & 1.63 & 1.70$^{+0.10}_{-0.09}$ & 0.82 & 0.80$^{+0.26}_{-0.31}$ & 0.82 & 0.71$^{+0.30}_{-0.40}$ & 11.90 & 11.89$^{+0.06}_{-0.09}$ & 12.18 & 12.14$^{+0.03}_{-0.03}$ & 0.47 & 0.43$^{+0.13}_{-0.11}$ \\
		CE-27 & 1.40 & 1.71$^{+0.09}_{-0.11}$ & 0.90 & 1.05$^{+0.19}_{-0.28}$ & 0.81 & 1.01$^{+0.21}_{-0.34}$ & 11.56 & 11.94$^{+0.08}_{-0.17}$ & 12.01 & 12.21$^{+0.03}_{-0.03}$ & 0.64 & 0.47$^{+0.17}_{-0.13}$ \\
		CE-28 & 1.53 & 1.46$^{+0.05}_{-0.07}$ & 0.90 & 1.01$^{+0.15}_{-0.22}$ & 0.81 & 0.95$^{+0.17}_{-0.29}$ & 11.81 & 11.91$^{+0.08}_{-0.09}$ & 12.16 & 12.33$^{+0.03}_{-0.04}$ & 0.55 & 0.62$^{+0.09}_{-0.09}$ \\
		\hline
		\hline
	\end{tabular}
\end{table*}

\end{document}